\documentclass[fleqn,11pt]{article}
\usepackage{amscd,amsmath,amssymb,verbatim}
\usepackage{amsthm}
\usepackage[pdftex]{graphicx}
\usepackage{epstopdf}
\usepackage[T1]{fontenc}
\usepackage{ae,aecompl}
\usepackage{subfigure}
\usepackage{array}
\usepackage{mathrsfs}
\usepackage[nohead]{geometry}
\usepackage[singlespacing]{setspace}
\usepackage{rotating}
\usepackage{multirow}
\usepackage{threeparttable}
\usepackage{enumerate}
\usepackage{subfigure}
\usepackage{placeins}
\usepackage{color}
\usepackage{accents}
\usepackage{tabularx}%for tables width user-defined width
\usepackage{booktabs}
\usepackage[round]{natbib}
\usepackage{float}
\numberwithin{equation}{section}
\DeclareMathOperator{\plim}{plim}

\DeclareMathOperator{\argmin}{argmin} 
\DeclareMathOperator{\argmax}{argmax}
 
\newcommand{\bs}{\boldsymbol}
\newcommand{\E}{\mathbb{E}}
\newcommand{\tr}{\mathrm{tr}}

\newcommand{\mf}{\mathbf}

\newcommand{\xvec}{\boldsymbol}

\DeclareMathOperator{\e}{\varepsilon}
\usepackage{hyperref}
\makeatletter
\newcommand{\distas}[1]{\mathbin{\overset{#1}{\kern\z@\sim}}}%
\newsavebox{\mybox}\newsavebox{\mysim}
\newcommand{\distras}[1]{%
  \savebox{\mybox}{\hbox{\kern3pt$\scriptstyle#1$\kern3pt}}%
  \savebox{\mysim}{\hbox{$\sim$}}%
  \mathbin{\overset{#1}{\kern\z@\resizebox{\wd\mybox}{\ht\mysim}{$\sim$}}}%
}
\makeatother

\renewcommand{\hat}[1]{\widehat{\text{$#1$}}}

\makeatletter
\newsavebox\myboxA
\newsavebox\myboxB
\newlength\mylenA

\renewcommand*\bar[2][0.5]{%
    \sbox{\myboxA}{$\m@th#2$}%
    \setbox\myboxB\null% Phantom box
    \ht\myboxB=\ht\myboxA%
    \dp\myboxB=\dp\myboxA%
    \wd\myboxB=#1\wd\myboxA% Scale phantom
    \sbox\myboxB{$\m@th\overline{\copy\myboxB}$}%  Overlined phantom
    \setlength\mylenA{\the\wd\myboxA}%   calc width diff
    \addtolength\mylenA{-\the\wd\myboxB}%
    \ifdim\wd\myboxB<\wd\myboxA%
       \rlap{\hskip 0.5\mylenA\usebox\myboxB}{\usebox\myboxA}%
    \else
        \hskip -0.5\mylenA\rlap{\usebox\myboxA}{\hskip 0.5\mylenA\usebox\myboxB}%
    \fi}
\makeatother
\hoffset 
\voffset

\author{Philipp Otto\thanks{School of Mathematics and Statistics, University of Glasgow, United Kingdom, email: philipp.otto@glasgow.ac.uk.} \and Osman Do\u{g}an\thanks{Department of Economics, Istanbul Technical University, Istanbul, T{\" u}rkiye, email: osmandogan@itu.edu.tr.} \and S\"uleyman Ta\c{s}p{\i}nar\thanks{Department of Economics, Queens College, The City University of New York, United States, email: staspinar@qc.cuny.edu.}}\medskip

\date{\today}
\title{Dynamic Spatiotemporal ARCH Models: Small and Large Sample Results}

\begin{document}
\maketitle
\sloppy

\singlespacing

\begin{abstract}
\noindent 
 This paper explores the estimation of a dynamic spatiotemporal autoregressive conditional heteroscedasticity (ARCH) model. The log-volatility term in this model can depend on (i) the spatial lag of the log-squared outcome variable, (ii) the time-lag of the log-squared outcome variable, (iii) the spatiotemporal lag of the log-squared outcome variable, (iv) exogenous variables, and (v) the unobserved heterogeneity across regions and time, i.e., the regional and time fixed effects. We examine the small and large sample properties of two quasi-maximum likelihood estimators and a generalized method of moments estimator for this model. We first summarize the theoretical properties of these estimators and then compare their finite sample properties through Monte Carlo simulations.
\end{abstract}
\vspace{7cm}
\noindent
JEL-Classification: C11, C23, C58.\\
Keywords:  Spatial ARCH, GMM, QMLE, volatility clustering, volatility, spatial dependence.
\newpage
\onehalfspacing
%\doublespacing
%\singlespacing

\section{Introduction}\label{intro}
This paper investigates the small and large sample properties of three estimators for dynamic spatiotemporal ARCH models suggested by \citet{Otto:2023}. This model allows the log-volatility term to depend on (i) the spatial lag of the log-squared outcome variable, (ii) the time-lag of the log-squared outcome variable, (iii) the spatiotemporal lag of the log-squared outcome variable, (iv) exogenous variables, and (v) the unobserved heterogeneity across regions and time, i.e., the regional and time fixed effects.  The estimation equation of the model is obtained through a log-squared transformation \citep{Robinson:2009, Taspinar:2021}.  Notably, while the estimation equation obtained is in the form of a standard spatial dynamic panel data model considered by \citet{Yu:2008} and \citet{Lee:2010}, it incorporates two important new features.

Firstly, the outcome variable, the spatial, temporal, and spatiotemporal lags are formulated using the log-squared original outcome variable. Secondly, the disturbance term in the model is the log-square of the original disturbance term due to the log-squared transformation. Following \citet{Lee:2007} and \citet{Lee:2014}, \citet{Otto:2023} consider a generalized method of moment (GMM) estimator based on a set of linear and quadratic moment functions. In this paper, we also consider two quasi-maximum likelihood (QML) estimators considered in \citet{Yu:2008} and \citet{Lee:2010} for the estimation of the model. The first estimator is based on a transformation approach requiring the estimation of regional fixed effects along with the other model parameters. The second estimator is based on a direct approach and necessitates the estimation of both regional and time fixed effects. We first compare the theoretical properties of these estimators and subsequently investigate their small-sample properties through Monte Carlo simulations.

The rest of the paper proceeds in the following way. In Section~\ref{sec3}, we describe the dynamic spatiotemporal ARCH model and show how it differs from a standard spatial dynamic panel data model. In Section~\ref{sec3}, we summarize the QML and the GMM estimation approaches and provide their large sample results. In Section~\ref{sec4}, we investigate the finite sample properties of both estimation approaches through Monte Carlo simulations.  In Section~\ref{sec5}, we conclude and provide some directions for future studies.

\section{Dynamic spatiotemporal ARCH model}\label{sec2}
We consider the random process $\{Y_t(\mf{s}): \mf{s}\in \mf{D}_1\subseteq\mathbb{R}^d, d>1, t\in D_2\subseteq\mathbb{R}\}$, where $\mf{s}\in \mf{D}_1$ denotes the spatial location, and $t\in D_2$ is the time point. The structure of the spatial domain 
$\mf{D}_1$ and the time domain $D_2$ depends on the nature of spatial data, and we will assume that $\mf{D}_1=\{\mf{s}_1,\hdots,\mf{s}_n\}$ and $D_2=\{1,2,\hdots,T\}$.   Then, the dynamic spatiotemporal ARCH model suggested by \citet{Otto:2023} can be expressed as 
\begin{align}
&Y_{t}(\mf{s}_i)=h_{t}^{1/2}(\mf{s}_i)\e_{t}(\mf{s}_i),\label{2.1}\\
&\log h_{t}(\mf{s}_i)=\sum_{j=1}^n\rho_{0}m_{ij}\log Y^2_{t}(\mf{s}_j)+\gamma_0\log Y^2_{t-1}(\mf{s}_j)+\sum_{j=1}^n\delta_{0}m_{ij}\log Y^2_{t-1}(\mf{s}_j)\label{2.2}\nonumber\\ 
&\quad\quad\quad\quad\quad+\mathbf{x}^{'}_{t}(\mf{s}_i)\bs{\beta}_0+\mu_{0}(\mf{s}_i)+\alpha_{t0},
\end{align}
where $h_{t}(\mf{s}_i)$ is considered as the volatility term in location $\mf{s}_i$ at time $t$, and $\e_{it}(\mf{s}_i)$ are independent and identically distributed random variables that have mean zero and unit variance. The log-volatility terms follow the process specified in \eqref{2.2},  where $\{m_{ij}\}$, for $i,j=1,\hdots,n$, are the non-stochastic spatial weights, $\mf{x}_{it}$ is a $k\times1$ vector of exogenous variables with the associated parameter vector $\bs{\beta}_0$, and  $\bs{\mu}_0=(\mu_{0}(\mf{s}_1),\hdots,\mu_{0}(\mf{s}_n))^{'}$ and $\bs{\alpha}_0=(\alpha_{10},\hdots,\alpha_{T0})^{'}$ are spatial and time fixed  effects. In the log-volatility equation, the spatial, temporal and spatiotemporal effects of the log-squared outcome variable on the log-volatility are measured by the unknown scalar parameters $\gamma_0$, $\rho_{0}$, and $\delta_{0}$, respectively. We assume that both $\bs{\mu}_0$ and $\bs{\alpha}_0$ can be correlated with the exogenous variables in an arbitrary manner and the initial value vector $\mf{Y}_0=(Y_{0}(\mf{s}_1),\hdots,Y_{0}(\mf{s}_n))^{'}$ is observable.

\begin{figure}
	\begin{center}
	    \includegraphics[width=0.6\textwidth]{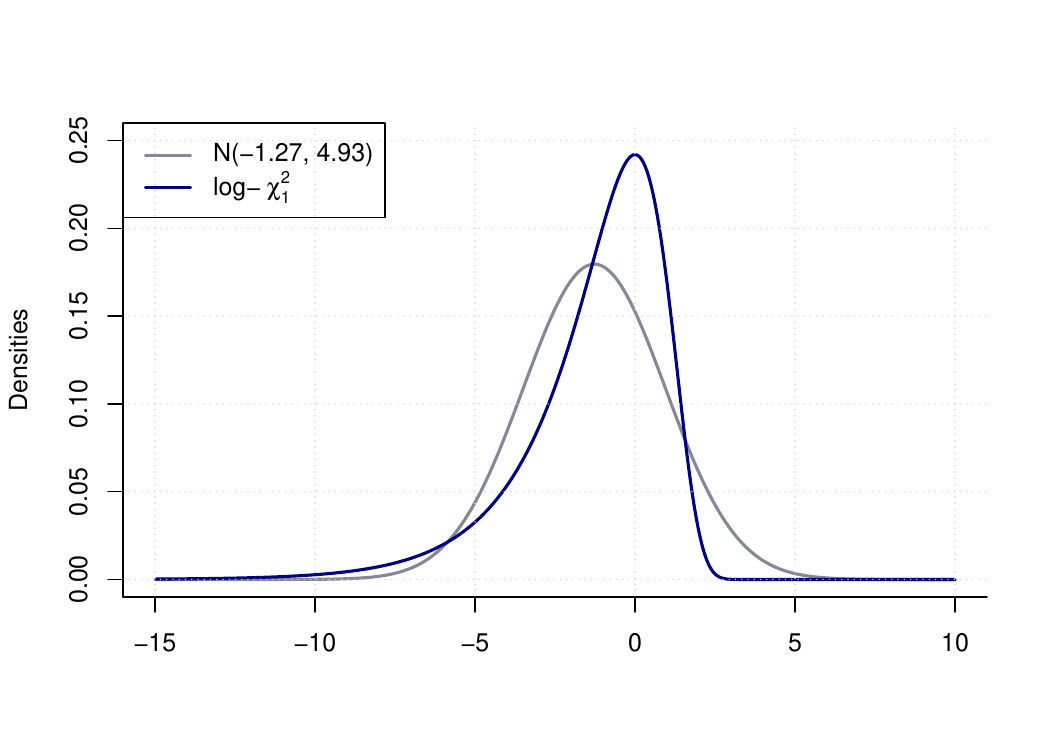}
	\end{center}
	\vspace{-0.80cm}
	\caption{The density plots of $N(-\- \log(2),\,\pi^2/2)$ and  $\log\chi^2_1$}
	\label{fig:logchi}
\end{figure}

Define $Y^*_{t}(\mf{s}_i)=\log Y^2_{t}(\mf{s}_i)$, $h^{*}_{t}(\mf{s}_i)=\log h_{t}(\mf{s}_i)$ and $\e^*_{t}(\mf{s}_i)=\log\e^2_{t}(\mf{s}_i)$. Then, we apply the log-squared transformation to \eqref{2.1} and obtain
\begin{align}\label{2.3}
Y^{*}_{t}(\mf{s}_i)=h^{*}_{t}(\mf{s}_i)+\e^{*}_{t}(\mf{s}_i),
\end{align}
In vector form, we can express \eqref{2.3} and \eqref{2.2} in the following way:
\begin{align}
&\mf{Y}^{*}_t=\mf{h}^{*}_t+\bs{\e}^{*}_t,\label{2.4}\\
&\mf{h}^{*}_t=\rho_{0}\mf{M}\mf{Y}^{*}_t+\gamma_0 \mf{Y}^{*}_{t-1}+\delta_{0}\mf{M}\mf{Y}^{*}_{t-1}+\mf{X}_t\bs{\beta}_0+\bs{\mu}_0+\alpha_{t0}\mf{1}_n,\label{2.5}
\end{align}
where $\mf{M}=(m_{ij})$ is the $n\times n$ matrix of the spatial weights, $\mf{Y}^{*}_t=(Y^{*}_{t}(\mf{s}_1),\hdots,Y^{*}_{t}(\mf{s}_n))^{'}$, $\mf{h}^{*}_t=(h^{*}_{t}(\mf{s}_1),\hdots,h^{*}_{t}(\mf{s}_n))^{'}$, $\bs{\e}^{*}_t=(\e^{*}_{t}(\mf{s}_1),\hdots,\e^{*}_{t}(\mf{s}_n))^{'}$, $\mf{X}_t=(\mf{x}_{t}(\mf{s}_1),\hdots,\mf{x}_{t}(\mf{s}_n))^{'}$, and $\mf{1}_n$ is the $n\times1$ vector of ones.  Then, we obtain the following estimation equation by substituting \eqref{2.5} into \eqref{2.4}: 
\begin{align}\label{2.6}
\mf{Y}^{*}_t=\rho_{0}\mf{M}\mf{Y}^{*}_t+\gamma_0 \mf{Y}^{*}_{t-1}+\delta_{0}\mf{M}\mf{Y}^{*}_{t-1}+\mf{X}_t\bs{\beta}_0+\bs{\mu}_0+\alpha_{t0}\mf{1}_n+\bs{\e}^{*}_t.
\end{align}
The estimation equation in \eqref{2.2} is in the form of a spatial dynamic panel data model considered by \citet{Yu:2008} and \citet{Lee:2010}.  However, it differs from a standard spatial dynamic panel data model in two important ways, which have implications for the chosen estimation approach. Firstly, the outcome variable, the spatial lag term ($\mf{M}\mf{Y}^{*}_t$),  and the spatiotemporal lag term ($\mf{M}\mf{Y}^{*}_{t-1}$)  are formulated in terms of the log-squared outcome variable. Secondly, the elements of $\bs{\e}^{*}_t$ are the log-squared original disturbance terms, i.e., $\e^*_{t}(\mf{s}_i)=\log\e^2_{t}(\mf{s}_i)$. If we assume that $\e_{t}(\mf{s}_i)\sim N(0,1)$, then $\e^{*}_{t}(\mf{s}_i)\sim\log\chi^2_1$, which is the log-chi squared distribution with one degree of freedom with the following density:
\begin{align}\label{2.7} 
p(\e^{*}_{t}(\mf{s}_i))=\frac{1}{\sqrt{2\pi}}\exp\left(-\frac{1}{2}\left(\exp(\e^{*}_{t}(\mf{s}_i))-\e^{*}_{t}(\mf{s}_i)\right)\right),\quad-\infty<\e^{*}_{t}(\mf{s}_i)<\infty.
\end{align}
The first two moments of $\log\chi^2_1$ are $\E(\e^{*}_{t}(\mf{s}_i)) = -\- \log(2) \approx-1.2704$, where $\gamma$ is Euler's constant, and $\text{Var}(\e^{*}_{t}(\mf{s}_i))=\pi^2/2\approx4.9348$ \citep[pp. 379-380]{Peter:2012}. In Figure~\ref{fig:logchi}, we compare this density with that of $N(-\- \log(2),\,\pi^2/2)$. As seen from the figure, the log-chi squared distribution exhibits significant skewness with a long left tail. 

%%%
\section{Estimation Approaches}\label{sec3}
Although the elements of $\bs{\e}^{*}_t$ in \eqref{2.6} are i.i.d across $\mf{s}_i$ and $t$, we may have $\E\left(\bs{\e}^{*}_t\right)\ne\mf{0}$ because of the log-squared transformation. Let $\E\left(\bs{\e}^{*}_t\right)=\mu_{\e}\mf{1}_n$, where $\mu_{\e}$ is a scalar unknown parameter, and  define $\mf{U}_t=(u_{t}(\mf{s}_1),\hdots,u_{t}(\mf{s}_n))^{'}=\bs{\e}^{*}_t-\E\left(\bs{\e}^{*}_t\right)=\bs{\e}^{*}_t-\mu_{\e}\mf{1}_n$. Then, we can express the estimation equation as
\begin{align}\label{3.1}
\mf{Y}^{*}_t&=\rho_{0}\mf{M}\mf{Y}^{*}_t+\gamma_0 \mf{Y}^{*}_{t-1}+\delta_{0}\mf{M}\mf{Y}^{*}_{t-1}+\mf{X}_t\bs{\beta}_0+\bs{\mu}_0 +\alpha_{t0}\mf{1}_n+\mu_{\e}\mf{1}_n+\mf{U}_t.
\end{align}
Following the spatial econometric literature, we consider two estimation approaches for  \eqref{3.1}: (i) the QML method, and (ii) the GMM method. In both approaches, we assume that $\{u_{t}(\mf{s}_i)\}$ are i.i.d across $t$ and $\mf{s}_i$ with mean zero and variance $\sigma^2_0$. 

\subsection{The QML Methods}\label{sec3.1}
Following \citet{Lee:2010}, we consider two QML methods: (i) the transformation approach and (ii) the direct approach. In the case of the transformation approach, we will eliminate the time fixed effects from the model through a suitable transformation and then estimate the regional fixed effects along with the other parameters. In the direct approach, we will estimate both $\bs{\alpha}_0$ and $\mf{c}_{n0}=\bs{\mu}_0+\mu_{\e}\mf{1}_n$ simultaneously. The transformation approach is only applicable if $\mf{M}$ is row-normalized, i.e., if $\mf{M}\mf{1}_n=\mf{1}_n$, while the direct approach does not require the row-normalization.

We start with the transformation approach and assume that $\mf{M}\mf{1}_n=\mf{1}_n$.  Let $\mf{J}_n=\mf{I}_n-\frac{1}{n}\mf{1}_n\mf{1}^{'}_n$, where $\mf{I}_n$ is the $n\times n$ the identity matrix, and $(\mf{F}_{n,n-1},\mf{1}_n/\sqrt{n})$ be the orthonormal matrix of eigenvectors of $\mf{J}_n$, where  $\mf{F}_{n,n-1}$ is the $n\times (n-1)$ matrix of eigenvectors corresponding to the eigenvalues of ones.\footnote{Some properties of $\mf{F}_{n,n-1}$ are $\mf{J}_n\mf{F}_{n,n-1}=\mf{F}_{n,n-1}$, $\mf{F}^{'}_{n,n-1}\mf{F}_{n,n-1}=\mf{I}_{n-1}$, $\mf{F}^{'}_{n,n-1}\mf{1}_{n}=\mf{0}$, $\mf{F}_{n,n-1}\mf{F}^{'}_{n,n-1}+\frac{1}{n}\mf{1}_n\mf{1}^{'}_n=\mf{I}_n$ and $\mf{F}_{n,n-1}\mf{F}^{'}_{n,n-1}=\mf{I}_n$. } Then, since $\mf{F}^{'}_{n,n-1}\mf{1}_n=\mf{0}$ holds, multiplying both sides of \eqref{3.1} with $\mf{F}^{'}_{n,n-1}$ yields 
\begin{align}\label{3.2}
\mf{Y}^{**}_t&=\rho_{0}\mf{M}^{*}\mf{Y}^{**}_t+\gamma_0 \mf{Y}^{**}_{t-1}+\delta_{0}\mf{M}^{*}\mf{Y}^{**}_{t-1}+\mf{X}^{*}_t\bs{\beta}_0+\bs{\mu}^{*}_0 +\mf{U}^{*}_t,
\end{align}
where $\mf{Y}^{**}_t=\mf{F}^{'}_{n,n-1}\mf{Y}^{*}_t$, $\mf{M}^{*}=\mf{F}^{'}_{n,n-1}\mf{M}\mf{F}_{n,n-1}$, $\mf{X}^{*}_t=\mf{F}^{'}_{n,n-1}\mf{X}_t$, $\bs{\mu}^{*}_0=\mf{F}^{'}_{n,n-1}\bs{\mu}_0$ and $\mf{U}^{*}_t=\mf{F}^{'}_{n,n-1}\mf{U}_t$.  Let $\bs{\theta}_0=(\gamma_0,\delta_0,\bs{\beta}^{'}_0,\rho_0,\sigma^2_0)^{'}$ denote the true parameter vector and $\bs{\theta}=(\gamma,\delta,\bs{\beta}^{'},\rho,\sigma^2)^{'}$ denote any other arbitrary value. If we assume that $\mf{U}_t\sim N(\mf{0},\sigma^2_0\mf{I}_n)$, then we have $\mf{U}^{*}_t\sim N(\mf{0},\sigma^2_0\mf{I}_{n-1})$. Thus, the log-likelihood function of \eqref{3.2} can be expressed as 
\begin{align}\label{3.3}
\ln L(\bs{\theta},\bs{\mu}^{*})&=-\frac{(n-1)T}{2}\ln(2\pi)-\frac{(n-1)T}{2}\ln(\sigma^2)+T\ln|\mf{I}_{n-1}-\rho\mf{M}^{*}|\nonumber\\
&-\frac{1}{2\sigma^2}\sum_{t=1}^T\mf{U}^{*'}_t(\bs{\theta})\mf{U}^{*}_t(\bs{\theta}),
\end{align}
where $\mf{U}^{*}_t(\bs{\theta})=(\mf{I}_{n-1}-\rho\mf{M}^{*})\mf{Y}^{**}_{t}-\mf{Z}^{**}_t\bs{\eta}-\bs{\mu}^{*}$ with $\mf{Z}^{**}_t=(\mf{Y}^{**}_{t-1},\mf{M}^{*}\mf{Y}^{**}_{t-1},\mf{X}^{*}_t)$ and $\bs{\eta}=(\gamma,\delta,\bs{\beta}^{'})^{'}$. Using the properties of $\mf{F}_{n,n-1}$, we can show that (i) $\ln|\mf{I}_{n-1}-\rho\mf{M}^{*}|=1/(1-\rho)|\mf{I}_n-\rho\mf{M}|$, and (ii) $\sum_{t=1}^T\mf{U}^{*'}_t(\bs{\theta})\mf{U}^{*}_t(\bs{\theta})=\sum_{t=1}^T\mf{U}^{'}_t(\bs{\theta})\mf{J}_n\mf{U}_t(\bs{\theta})$, where $\mf{U}_t(\bs{\theta})=(\mf{I}_{n-1}-\rho\mf{M})\mf{Y}^{*}_{t}-\mf{Z}^{*}_t\bs{\eta}-\bs{\mu}$ with   $\mf{Z}^{*}_t=(\mf{Y}^{*}_{t-1},\mf{M}\mf{Y}^{*}_{t-1},\mf{X}_t)$. Thus, we can express \eqref{3.3} in terms of the original variables as
\begin{align}\label{3.4}
\ln L(\bs{\theta},\bs{\mu})&=-\frac{(n-1)T}{2}\ln(2\pi)-\frac{(n-1)T}{2}\ln(\sigma^2)+T\ln(1-\rho)+T\ln|\mf{I}_{n}-\rho\mf{M}|\nonumber\\
&-\frac{1}{2\sigma^2}\sum_{t=1}^T\mf{U}^{'}_t(\bs{\theta})\mf{J}_n\mf{U}_t(\bs{\theta}).
\end{align}
For an $n\times1$ vector $\mf{V}_t$, we define $\tilde{\mf{V}}_t=\mf{V}_t-\frac{1}{T}\sum_{t=1}^T\mf{V}_t$. Then, concentrating out $\bs{\mu}$ from \eqref{3.4} yields
\begin{align}\label{3.5}
\ln L(\bs{\theta})&=-\frac{(n-1)T}{2}\ln(2\pi)-\frac{(n-1)T}{2}\ln(\sigma^2)+T\ln(1-\rho)+T\ln|\mf{I}_{n}-\rho\mf{M}|\nonumber\\
&-\frac{1}{2\sigma^2}\sum_{t=1}^T\tilde{\mf{U}}^{'}_t(\bs{\theta})\mf{J}_n\tilde{\mf{U}}_t(\bs{\theta}),
\end{align}
where $\tilde{\mf{U}}_t(\bs{\theta})=(\mf{I}_{n-1}-\rho\mf{M})\tilde{\mf{Y}}^{*}_{t}-\tilde{\mf{Z}}^{*}_t\bs{\eta}-\bs{\mu}$ with $\tilde{\mf{Z}}^{*}_t=(\tilde{\mf{Y}}^{*}_{t-1},\mf{M}\tilde{\mf{Y}}^{*}_{t-1},\tilde{\mf{X}}_t)$. Then, the QMLE $\hat{\bs{\theta}}_{nT}$ of $\bs{\theta}_0$ is defined by $\hat{\bs{\theta}}_{nT}=\argmax_{\bs{\theta}}\ln L(\bs{\theta})$. To investigate the asymptotic properties, following \citet{Lee:2010}, we can show that the score functions can be decomposed as $\frac{1}{\sqrt{(n-1)T}}\frac{\partial\ln L(\bs{\theta}_0)}{\partial\bs{\theta}}=\frac{1}{\sqrt{(n-1)T}}\frac{\partial\ln L^u(\bs{\theta}_0)}{\partial\bs{\theta}}-\Delta_{nT}$, where the first component is uncorrelated with $\mf{U}_t$ while the second component is correlated with $\mf{U}_t$ for $t\leq T-1$. Moreover, the second component is the source of the asymptotic bias with the order $\Delta_{nT}=\sqrt{\frac{n-1}{T}}\mf{a}+O(\sqrt{(n-1)/T^3})+O_p(1/T^3)$, where $\mf{a}=O(1)$. Then, following \citet{Lee:2010}, it can be shown that 
\begin{align}\label{3.6}
&\sqrt{T(n-1)}\left(\hat{\bs{\theta}}_{nT}-\bs{\theta}_0\right)+\sqrt{\frac{n-1}{T}}\mf{b}+O_p\left(\max\left\{\sqrt{(n-1)/T^3},\sqrt{1/T}\right\}\right)\nonumber\\
&\xrightarrow{d}N\left(\mf{0},\,\bs{\Sigma}^{-1}(\bs{\Sigma}+\bs{\Omega})\bs{\Sigma}^{-1}\right),
\end{align}
 where $\bs{\Sigma}=\lim_{T\to\infty}\E\left(-\frac{1}{(n-1)T}\frac{\partial^2\ln L(\bs{\theta}_0)}{\partial\bs{\theta}\partial\bs{\theta}^{'}}\right)$, $\bs{\Omega}=\lim_{T\to\infty}\text{Var}\left(\frac{1}{\sqrt{(n-1)T}}\frac{\partial\ln L^u(\bs{\theta}_0)}{\partial\bs{\theta}}\right)-\bs{\Sigma}$, and $\mf{b}=\bs{\Sigma}^{-1}\mf{a}=O(1)$ is the asymptotic bias term.\footnote{The explicit forms of $\mf{b}$, $\bs{\Sigma}$ and $\bs{\Omega}$ can be readily determined using the results presented in \citet{Lee:2010}. } According to the asymptotic result in \eqref{3.6}, there are three cases depending on the relative growth rates of $n$ and $T$. Firstly, if $\frac{n}{T}\to\infty$, i.e., if $n$ grows faster than $T$, then $T(\hat{\bs{\theta}}_{nT}-\bs{\theta}_0)+\mf{b}\xrightarrow{p}\mf{0}$. That is, $\hat{\bs{\theta}}_{nT}$ is consistent with rate $T$ and has a degenerate limiting distribution.  Secondly, if $\frac{n}{T}\to0$, then we will have $\sqrt{T(n-1)}(\hat{\bs{\theta}}_{nT}-\bs{\theta}_0)\xrightarrow{d}N\left(\mf{0},\,\bs{\Sigma}^{-1}(\bs{\Sigma}+\bs{\Omega})\bs{\Sigma}^{-1}\right)$. Lastly, when $\frac{n}{T}\to c\in \mathbb{R}_+$, i.e., when $T$ is asymptotically proportional to $n$, we have $\sqrt{T(n-1)}(\hat{\bs{\theta}}_{nT}-\bs{\theta}_0)+\sqrt{c}\mf{b}\xrightarrow{d}N\left(\mf{0},\,\bs{\Sigma}^{-1}(\bs{\Sigma}+\bs{\Omega})\bs{\Sigma}^{-1}\right)$. In this case, \citet{Lee:2010} suggest a bias corrected estimator defined  by $\hat{\bs{\theta}}^1_{nT}=\hat{\bs{\theta}}_{nT}-\hat{\mf{b}}$, where $\hat{\mf{b}}$ is a plug-in estimator of $\mf{b}$. Then, it follows that $\sqrt{T(n-1)}(\hat{\bs{\theta}}^1_{nT}-\bs{\theta}_0)\xrightarrow{d}N\left(\mf{0},\,\bs{\Sigma}^{-1}(\bs{\Sigma}+\bs{\Omega})\bs{\Sigma}^{-1}\right)$ when $\frac{n}{T^3}\to0$, i.e., when $T^3$ grows faster than $n$.

Next, we introduce the direct approach for the estimation of \eqref{3.1}. Under the assumption that $\mf{U}_t\sim N(\mf{0},\sigma^2_0\mf{I}_n)$, we can derive the log-likelihood function of \eqref{3.1} as
\begin{align}\label{3.7}
\ln L^d(\bs{\theta},\mf{c}_n,\bs{\alpha})&=-\frac{nT}{2}\ln(2\pi)-\frac{nT}{2}\ln(\sigma^2)+T\ln|\mf{I}_{n}-\rho\mf{M}|\nonumber\\
&-\frac{1}{2\sigma^2}\sum_{t=1}^T\mf{U}^{'}_t(\bs{\theta},\mf{c}_n,\bs{\alpha})\mf{U}_t(\bs{\theta},\mf{c}_n,\bs{\alpha}),
\end{align}
where $\mf{U}_t(\bs{\theta},\mf{c}_n,\bs{\alpha})=(\mf{I}_{n}-\rho\mf{M})\mf{Y}^{*}_{t}-\mf{Z}^{*}_t\bs{\eta}-\mf{c}_n+\alpha_t\mf{1}_n$. Concentrating both $\mf{c}_n$ and $\bs{\alpha}$ from \eqref{3.7} yields
\begin{align}\label{3.8}
\ln L^d(\bs{\theta})&=-\frac{nT}{2}\ln(2\pi)-\frac{nT}{2}\ln(\sigma^2)+T\ln|\mf{I}_{n}-\rho\mf{M}|-\frac{1}{2\sigma^2}\sum_{t=1}^T\tilde{\mf{U}}^{'}_t(\bs{\theta})\mf{J}_n\tilde{\mf{U}}_t(\bs{\theta}). 
\end{align}
 The QMLE $\hat{\bs{\theta}}^d_{nT}$ of $\bs{\theta}_0$ is defined by $\hat{\bs{\theta}}^d_{nT}=\argmax_{\bs{\theta}}\ln L^d(\bs{\theta})$. As in the case of the transformation approach, the score functions can be decomposed as $\frac{1}{\sqrt{nT}}\frac{\partial \ln L^d(\bs{\theta})}{\partial\bs{\theta}}=\frac{1}{\sqrt{nT}}\frac{\partial \ln L^{du}(\bs{\theta})}{\partial\bs{\theta}}-\Delta_{1,nT}-\Delta_{2,nT}$, where the first component is uncorrelated with $\mf{U}_t$ and the other components are the sources of asymptotic bias with the following orders: $\Delta_{1,nT}=\sqrt{\frac{n}{T}}\mf{a}_1+O(\sqrt{n/T^3})+O_p(1/\sqrt{T})$ with $\mf{a}_1=O(1)$, and  $\Delta_{2,nT}=\sqrt{\frac{n}{T}}\mf{a}_2$ with $\mf{a}_2=O(1)$. Then, it follows from \citet{Lee:2010} that
\begin{align}\label{3.9}
&\sqrt{nT)}\left(\hat{\bs{\theta}}^d_{nT}-\bs{\theta}_0\right)+\sqrt{\frac{n}{T}}\mf{b}_1+\sqrt{\frac{T}{n}}\mf{b}_2+O_p\left(\max\left\{\sqrt{n/T^3},\sqrt{T/n^3},\sqrt{1/T}\right\}\right)\nonumber\\
&\xrightarrow{d}N\left(\mf{0},\,\bs{\Sigma}^{-1}(\bs{\Sigma}+\bs{\Omega})\bs{\Sigma}^{-1}\right),
\end{align}
 where $\bs{\Sigma}=\lim_{T\to\infty}\E\left(-\frac{1}{nT}\frac{\partial^2\ln L^d(\bs{\theta}_0)}{\partial\bs{\theta}\partial\bs{\theta}^{'}}\right)$, $\bs{\Omega}=\lim_{T\to\infty}\text{Var}\left(\frac{1}{\sqrt{nT}}\frac{\partial\ln L^{du}(\bs{\theta}_0)}{\partial\bs{\theta}}\right)-\bs{\Sigma}$, and $\mf{b}_1=\bs{\Sigma}^{-1}\mf{a}_1=O(1)$ and $\mf{b}_2=\bs{\Sigma}^{-1}\mf{a}_2=O(1)$ are the asymptotic bias terms. As in the transformation case, there are three cases depending on the relative growth rates of $n$ and $T$. The first two cases are the degenerate limiting distribution cases, occurring when either $\frac{n}{T}\to0$ or $\frac{n}{T}\to\infty$ holds. If $\frac{n}{T}\to0$ holds, then we have $n(\hat{\bs{\theta}}^d_{nT}-\bs{\theta}_0)+\mf{b}_2\xrightarrow{p}\mf{0}$, and when $\frac{n}{T}\to\infty$ holds, we have $T(\hat{\bs{\theta}}^d_{nT}-\bs{\theta}_0)+\mf{b}_1\xrightarrow{p}\mf{0}$. Finally, when $\frac{n}{T}\to c\in \mathbb{R}_+$ holds, we have $\sqrt{nT)}(\hat{\bs{\theta}}^d_{nT}-\bs{\theta}_0)+\sqrt{c}\mf{b}_1+\sqrt{1/c}\mf{b}_2\xrightarrow{d}N\left(\mf{0},\,\bs{\Sigma}^{-1}(\bs{\Sigma}+\bs{\Omega})\bs{\Sigma}^{-1}\right)$, which indicates that we need a bias correction for $\hat{\bs{\theta}}^d_{nT}$. Let $\bs{\theta}^{d1}_{nT}=\hat{\bs{\theta}}^d_{nT}-\hat{\mf{b}}_1/T-\hat{\mf{b}}_2/T$ be the bias corrected estimator, where $\hat{\mf{b}}_1$ and $\hat{\mf{b}}_2$ are the plug-in estimators of $\mf{b}_1$ and $\mf{b}_2$, respectively. Then, it can be shown that if  $\frac{n}{T^3}\to0$ and $\frac{T}{n^3}\to\infty$, then we have $\sqrt{nT)}(\hat{\bs{\theta}}^{d1}_{nT}-\bs{\theta}_0)\xrightarrow{d}N\left(\mf{0},\,\bs{\Sigma}^{-1}(\bs{\Sigma}+\bs{\Omega})\bs{\Sigma}^{-1}\right)$.

\subsection{The GMM Method}\label{sec3.2}
In the GMM approach, we will use two different transformations to eliminate $\bs{\mu}$, $\bs{\alpha}$ and $\mu_{\e}$ from the model. The first transformation is based on the decomposition of $\mf{J}_T=\left(\mf{I}_T-\frac{1}{T}\mf{1}_T\mf{1}^{'}_T\right)$, where $\mf{I}_T$ is the $T\times T$ identity matrix. Let $\left(\mf{F}_{T,T-1},\mf{1}_T/\sqrt{T}\right)$ be the orthonormal  eigenvector matrix of $\mf{J}_T$, where $\mf{F}_{T,T-1}$ is the $T\times (T-1)$ sub-matrix  containing eigenvectors corresponding to the eigenvalues of one.  Let  $\mf{D}=\left(\mf{d}_1,\hdots,\mf{d}_T\right)$ be an $n\times T$ matrix, where $\mf{d}_t$ is an $n\times1$ column vector for $t=1,\hdots,T$.  Using $\mf{F}_{T,T-1}$, we can transform $\mf{D}$ into an $n\times(T-1)$ matrix in the following way: $\mf{D}^{*}=\left(\mf{d}^{*}_1,\hdots,\mf{d}^{*}_{T-1}\right)=\left(\mf{d}_1,\hdots,\mf{d}_T\right)\mf{F}_{T,T-1}$. Since $\left(\bs{\mu}_0,\hdots,\bs{\mu}_0\right)\mf{F}_{T,T-1}=\bs{\mu}_0\mf{1}^{'}_T\mf{F}_{T,T-1}=\mf{0}_{n\times(T-1)}$ and $\left(\mu_{\e}\mf{1}_n,\hdots,\mu_{\e}\mf{1}_n\right)\mf{F}_{T,T-1}=\mu_{\e}\mf{1}_n\mf{1}^{'}_T\mf{F}_{T,T-1}=\mf{0}_{n\times(T-1)}$, this transformation can remove both $\bs{\mu}_0$ and $\mu_{\e}$ from the model.  If we apply  $\mf{F}_{T,T-1}$ to \eqref{3.1} in a similar manner, we will obtain
\begin{align}\label{3.10}
\mf{Y}^{**}_t&=\rho_{0}\mf{M}\mf{Y}^{**}_t+\_0\mf{Y}^{**,-1}_{t-1}+\delta_{0}\mf{M}\mf{Y}^{**,-1}_{t-1}+\mf{X}^{*}_t\bs{\beta}_0+\alpha^{*}_{t0}\mf{1}_n+\mf{U}^{*}_t.
\end{align}
 Note that \eqref{3.10} still includes the transformed time fixed effects denoted by $\alpha^{*}_{t0}$ for $t=1,\hdots,T$. To remove these effects, we apply a second transformation by pre-multiplying \eqref{3.10} with $\mf{J}_n$:
\begin{align}\label{3.11}
\mf{J}_n\mf{Y}^{**}_t&=\rho_{0}\mf{J}_n\mf{M}\mf{Y}^{**}_t+\_0\mf{J}_n\mf{Y}^{**,-1}_{t-1}+\delta_{0}\mf{J}_n\mf{M}\mf{Y}^{**,-1}_{t-1}+\mf{J}_n\mf{X}^{*}_t\bs{\beta}_0+\mf{J}_n\mf{U}^{*}_t.
\end{align}
Following \citet{Lee:2007} and \citet{Lee:2014}, \citet{Otto:2023} consider both linear and quadratic moment functions for the estimation of \eqref{3.11}. Let $N=n(T-1)$, $\bs{\theta}=(\rho,\bs{\eta}^{'})^{'}$ and $\mf{J}_N=\mf{I}_{T-1}\otimes\mf{J}_n$. Then, the vector of moment functions takes the following form: 
\begin{align}\label{3.12}
\mf{g}_N(\bs{\theta})=
\begin{pmatrix}
\mf{U}^{'}_N(\bs{\theta})\mf{J}_N\mf{P}_{1N}\mf{J}_N\mf{U}_N(\bs{\theta})\\
\vdots\\
\mf{U}^{'}_N(\bs{\theta})\mf{J}_N\mf{P}_{mN}\mf{J}_N\mf{U}_N(\bs{\theta})\\
\mf{Q}^{'}_N\mf{J}_N\mf{U}_N(\bs{\theta})
\end{pmatrix},
\end{align}
where $\mf{U}_N(\bs{\theta})=(\mf{U}^{*'}_1(\bs{\theta}),\hdots,\mf{U}^{*'}_{T-1}(\bs{\theta}))^{'}$ with $\mf{U}^{*}_t(\bs{\theta})=(\mf{I}_n-\rho\mf{M})\mf{Y}^{**}_t-\mf{Z}^{**}_t\bs{\eta}-\alpha^{*}_{t}\mf{1}_n$. In \eqref{3.12}, the linear moment function takes the form of $\mf{Q}^{'}_N\mf{J}_N\mf{U}_N(\bs{\theta})$, where $\mf{Q}_N$ is the  $n(T-1)\times k_q$ matrix of IVs, and the quadratic moment function takes the form  of $\mf{U}^{'}_N(\bs{\theta})\mf{J}_N\mf{P}_{jN}\mf{J}_N\mf{U}_N(\bs{\theta})$, where $\mf{P}_{jN}=\mf{I}_{T-1}\otimes\mf{P}_j$ and $\mf{P}_j$ is an $n\times n$ matrix satisfying $\tr(\mf{J}_n\mf{P}_j\mf{J}_n)=0$ for $j=1,2,\hdots,m$. 

Let $\bs{\Omega}_N=\frac{1}{N}\E\left(\mf{g}_N(\bs{\theta}_0)\mf{g}^{'}_N(\bs{\theta}_0)\right)$. Then, the optimal GMME is defined by
\begin{align}\label{3.13}
\hat{\bs{\theta}}_N=\argmin_{\bs{\theta}\in\bs{\Theta}}\mf{g}^{'}_N(\bs{\theta})\hat{\bs{\Omega}}^{-1}_N\mf{g}_N(\bs{\theta}),
\end{align}
where $\hat{\bs{\Omega}}_N$ be a consistent estimator of $\bs{\Omega}_N$, i.e., $\hat{\bs{\Omega}}_N-\bs{\Omega}_N=o_p(1)$. Under some assumptions, \citet{Otto:2023} show that $\frac{1}{N}\frac{\partial \mf{g}_N(\bs{\theta}_0)}{\partial\bs{\theta}^{'}}=\mf{D}_{1N}+\mf{D}_{2N}+O_p(N^{-1/2})$, where $\mf{D}_{1N}=O(1)$ and $\mf{D}_{2N}=O(T^{-1})$.\footnote{See \citet{Otto:2023}  for the explicit forms of $\mf{D}_{1N}$ and $\mf{D}_{2N}$.} Then, when $T$ is finite and $n\to\infty$, it can be shown that
\begin{align}\label{3.14}
\sqrt{n}\left(\hat{\bs{\theta}}_N-\bs{\theta}_0\right)\xrightarrow{d}N\left(\mf{0},\,\plim_{n\to\infty}\frac{1}{T-1}\left(\left(\mf{D}_{1N}+\mf{D}_{2N}\right)^{'}\bs{\Omega}^{-1}_N\left(\mf{D}_{1N}+\mf{D}_{2N}\right)\right)^{-1}\right).
\end{align}
On the other hand, when $T\to\infty$, we have $\mf{D}_{2N}=O(T^{-1})=o(1)$. Thus,   when $T\to\infty$ and $n\to\infty$, \citet{Otto:2023} show that
\begin{align}\label{3.15}
\sqrt{N}\left(\hat{\bs{\theta}}_N-\bs{\theta}_0\right)\xrightarrow{d}N\left(\mf{0},\,\plim_{n,T\to\infty}\left(\mf{D}^{'}_{1N}\bs{\Omega}^{-1}_N\mf{D}_{1N}\right)^{-1}\right).
\end{align}
Note that the asymptotic results in \eqref{3.14} and \eqref{3.15} hold for the arbitrary IV matrix $\mf{Q}_N$ and the quadratic moment matrices $\mf{P}_{jN}$ for $j=1,2,\hdots,m$. The asymptotic efficiency of $\hat{\bs{\theta}}_N$ should be considered in choosing the IV and quadratic moment matrices. Following \citet{Lee:2014}, \citet{Otto:2023} provide a vector of moment functions that can lead to the most efficient GMME when both $n$ and $T$ are large. 

%%%%%%%%%%%%%%%%%%%%%%%%
\section{Monte Carlo Simulations}\label{sec4}

In this section, we investigate the finite sample performance of all three considered estimators when the sample size is small. In particular, we are looking at the case of short time series (i.e., $T$ is small), which is often encountered in practice for geo-referenced data. More precisely, we simulate $Y_{t}(\xvec{s}_i) = h_{t}(\xvec{s}_i)^{1/2}\varepsilon_{t}(\xvec{s}_i)$ with 
\begin{equation}
\log h_{t}(\xvec{s}_i) = \sum_{j=1}^n\rho_{0}m_{ij}\log Y^2_{t}(\xvec{s}_j)+\gamma_0\log Y^2_{t-1}(\xvec{s}_i) + \sum_{j=1}^n\delta_{0}m_{ij}\log Y^2_{t-1}(\xvec{s}_j) + \mathbf{x}^{'}_{it}\bs{\beta}_0 + \mu_{i0} + \alpha_{t0}. 
\end{equation}
We consider three different data-generating processes with the following parameter settings: 
\begin{align*}
&M_1:\, \rho_0 = 0.2, \,\gamma_0 = 0.5, \, \delta_0 = -0.2, \, \xvec{\beta}_0 = (0.5, 1)', \alpha_{t0} = 0 \, \forall t, \\
&M_2:\, \rho_0 = 0.3, \, \gamma_0 = 0.2, \, \delta_0 = 0.2, \, \xvec{\beta}_0 = (0.5, 1)', \alpha_{t0} = 0 \, \forall t,\\
&M_3:\, \rho_0 = 0.8, \gamma_0 = 0.1, \, \delta_0 = -0.2, \, \xvec{\beta}_0 = (0.5, 1)'.
\end{align*}
In $M_1$, the temporal effect is relatively dominant; in $M_2$, all effects are relatively weak; and finally, in $M_3$, the spatial effect is relatively strong. Both $M_1$ and $M_2$ include only the spatial fixed effects, while $M_3$ includes both the spatial and time fixed effects. The spatial/time fixed effects and the exogenous regressors are independently simulated from $N(0,1)$. We generate the error terms $\varepsilon_{t}(\xvec{s}_i)$'s independently from $N(0,1)$, and consider a queen contiguity row-normalized weights matrix. The sample size gradually increases for both $n$ and $T$ with $n \in \{25, 49, 81\}$ on a regular two-dimensional lattice (with side length 5, 7, and 9) and $T \in \{ 5, 10, 20\}$. The number of repetitions is set to $1000$ in all cases. Thus, we considered 27 different model specifications in total and three different estimators, which were always applied to the same simulated values.

%For each estimator, we report the root mean square errors (RMSE), the mean absolute errors (MAE), and the average bias (BIAS). The results are reported in Tables~\ref{table:rmse} and \ref{table:bias}.\footnote{The MAE results are provided in the accompanying web appendix. These results suggest the same conclusions as those obtained from the RMSE results presented in Table~\ref{table:rmse}. Additionally, in the accompanying web appendix, we offer a graphical representation of the results provided in Tables~\ref{table:rmse} to \ref{table:bias}.} In Table~\ref{table:rmse}, we observe that the RMSE decreases when either $n$ or $T$ increases for all estimators. In the case of $\rho$, the QMLE based on the direct approach is relatively more efficient than the other estimators.  For instance,  in $M_2$ with $(n,T)=(25,20)$, the RMSEs are $5.836$, $4.673$ and $1.981$ for the GMME, the QMLE based on transformation and the QMLE based on the direct approach, respectively. In the case of $\gamma$, the GMME is relatively more efficient than the QML estimators in all cases. In the case of $\delta$, the direct approach performs slightly better than the other approaches. All estimators perform similarly in the case of $\beta_0$ and $\beta_1$. 

For each estimator, we report bias (BIAS), the root mean square error (RMSE) and the mean absolute error (MAE). The results are reported in Tables~\ref{table:bias} and \ref{table:rmse}.\footnote{The MAE results are provided in the accompanying web appendix. These results suggest the same conclusions as those obtained from the RMSE results presented in Table~\ref{table:rmse}. Additionally, in the accompanying web appendix, we provide graphical representations of the results provided in Tables~\ref{table:bias} and \ref{table:rmse}.}

In Table~\ref{table:bias}, we observe that all estimators report relatively larger bias when $(n,T)=(25, 5)$, especially in the case of $\rho$, $\gamma$ and $\delta$.  When either $n$ or $T$ increases, all estimators impose a smaller bias in all cases. In the case of $\rho$, the QMLE based on the direct approach performs relatively better than other estimators. In the case of $\gamma$, the GMME performs better than the QML estimators. It is clear that the QML methods require a large $T$ to estimate this parameter accurately. In the case of $\delta$, the GMME seems to impose a relatively smaller bias than the QML estimators. For $\beta_0$ and $\beta_1$, the QMLE based on the transformation approach shows a significant bias when $T=5$. 
 
In Table~\ref{table:rmse}, we observe that the RMSE decreases when either $n$ or $T$ increases for all estimators. In the case of $\rho$, the QMLE based on the direct approach is relatively more efficient than the other estimators.  For instance,  in $M_2$ with $(n,T)=(25,20)$, the RMSEs are $0.185$, $0.148$ and $0.063$ for the GMME, the QMLE based on transformation and the QMLE based on the direct approach, respectively. In the case of $\gamma$, the GMME is relatively more efficient than the QML estimators in all cases. In the case of $\delta$, the direct approach performs slightly better than the other approaches. All estimators perform similarly in the case of $\beta_0$ and $\beta_1$.

% latex table generated in R 4.3.0 by xtable 1.8-4 package
% Wed Nov 15 18:10:43 2023
\begin{table}[H]
\centering
\caption{\footnotesize BIAS of the GMM and QML estimators across different parameter values and sample sizes} 
\label{table:bias}
\begin{scriptsize}
\begin{tabular}{lllrrrrrrrrr}
  \hline\hline
 &  &  &  & $n=25$ &  &  & $n=49$ &  &  & $n=81$ &  \\ 
Parameter & Model & Method & $T=5$ & $T=10$ & $T=20$ & $T=5$ & $T=10$ & $T=20$ & $T=5$ & $T=10$ & $T=20$ \\ 
  \hline
  &  & GMM & 0.273 & 0.120 & 0.057 & 0.237 & 0.115 & 0.049 & 0.163 & 0.071 & 0.035 \\ 
  & M1 & QML (transformed) & -0.155 & -0.134 & -0.125 & -0.074 & -0.070 & -0.065 & -0.043 & -0.041 & -0.039 \\ 
  &  & QML (direct) & -0.028 & -0.009 & -0.005 & -0.008 & -0.005 & -0.003 & -0.002 & -0.002 & -0.000 \\
  \cline{2-12}
  &  & GMM & 0.253 & 0.123 & 0.065 & 0.224 & 0.114 & 0.052 & 0.155 & 0.072 & 0.038 \\ 
  $\rho$ & M2 & QML (transformed) & -0.186 & -0.145 & -0.128 & -0.101 & -0.078 & -0.067 & -0.071 & -0.049 & -0.040 \\ 
  &  & QML (direct) & -0.079 & -0.026 & -0.010 & -0.052 & -0.019 & -0.006 & -0.046 & -0.016 & -0.004 \\ 
  \cline{2-12}
  &  & GMM & 0.180 & 0.057 & 0.024 & 0.141 & 0.061 & 0.025 & 0.097 & 0.043 & 0.022 \\ 
  & M3 & QML (transformed) & -0.153 & -0.129 & -0.122 & -0.063 & -0.060 & -0.054 & -0.035 & -0.035 & -0.032 \\ 
  &  & QML (direct) & -0.158 & -0.128 & -0.121 & -0.066 & -0.060 & -0.054 & -0.039 & -0.035 & -0.032 \\
    \hline
  &  & GMM & -0.101 & -0.007 & -0.000 & -0.056 & -0.004 & -0.001 & -0.033 & -0.002 & -0.000 \\ 
  & M1 & QML (transformed) & -0.404 & -0.165 & -0.074 & -0.394 & -0.160 & -0.072 & -0.389 & -0.157 & -0.071 \\ 
  &  & QML (direct) & -0.354 & -0.153 & -0.070 & -0.345 & -0.150 & -0.069 & -0.341 & -0.146 & -0.068 \\ 
  \cline{2-12}
  &  & GMM & -0.072 & -0.011 & -0.004 & -0.047 & -0.009 & -0.004 & -0.026 & -0.005 & -0.003 \\ 
  $\gamma$ & M2 & QML (transformed) & -0.319 & -0.128 & -0.055 & -0.315 & -0.127 & -0.057 & -0.312 & -0.126 & -0.058 \\ 
  &  & QML (direct) & -0.289 & -0.126 & -0.059 & -0.284 & -0.123 & -0.058 & -0.281 & -0.121 & -0.058 \\ 
   \cline{2-12}
  &  & GMM & -0.038 & -0.003 & -0.002 & -0.025 & -0.002 & -0.002 & -0.012 & -0.001 & -0.002 \\ 
  & M3 & QML (transformed) & -0.250 & -0.107 & -0.052 & -0.245 & -0.104 & -0.050 & -0.244 & -0.102 & -0.050 \\ 
  &  & QML (direct) & -0.233 & -0.103 & -0.051 & -0.230 & -0.100 & -0.049 & -0.228 & -0.098 & -0.049 \\
  \hline
  &  & GMM & -0.063 & -0.025 & -0.012 & -0.040 & -0.036 & -0.018 & -0.046 & -0.015 & -0.012 \\ 
  & M1 & QML (transformed) & 0.115 & 0.077 & 0.066 & 0.117 & 0.061 & 0.037 & 0.116 & 0.055 & 0.029 \\ 
  &  & QML (direct) & 0.091 & 0.041 & 0.020 & 0.102 & 0.039 & 0.014 & 0.098 & 0.041 & 0.016 \\ 
   \cline{2-12}
  &  & GMM & -0.143 & -0.048 & -0.017 & -0.093 & -0.050 & -0.020 & -0.090 & -0.026 & -0.012 \\ 
  $\delta$ & M2 & QML (transformed) & -0.021 & 0.025 & 0.045 & -0.011 & 0.014 & 0.020 & -0.012 & 0.008 & 0.014 \\ 
  &  & QML (direct) & -0.003 & -0.002 & 0.002 & 0.015 & -0.000 & -0.002 & 0.010 & 0.002 & 0.001 \\
   \cline{2-12} 
  &  & GMM & 0.063 & 0.012 & 0.009 & 0.059 & 0.016 & 0.006 & 0.035 & 0.015 & 0.007 \\ 
  & M3 & QML (transformed) & 0.194 & 0.073 & 0.032 & 0.214 & 0.081 & 0.031 & 0.217 & 0.085 & 0.036 \\ 
  &  & QML (direct) & 0.171 & 0.068 & 0.031 & 0.192 & 0.077 & 0.030 & 0.195 & 0.080 & 0.035 \\
  \hline
  &  & GMM & -0.023 & -0.004 & -0.009 & -0.017 & -0.006 & -0.003 & -0.016 & -0.004 & -0.001 \\ 
  & M1 & QML (transformed) & -0.139 & -0.059 & -0.033 & -0.136 & -0.058 & -0.027 & -0.138 & -0.060 & -0.026 \\ 
  &  & QML (direct) & -0.063 & -0.012 & -0.008 & -0.064 & -0.015 & -0.004 & -0.065 & -0.016 & -0.003 \\ 
   \cline{2-12}
  &  & GMM & -0.016 & -0.004 & -0.009 & -0.014 & -0.006 & -0.003 & -0.013 & -0.004 & -0.002 \\ 
  $\beta_0$ & M2 & QML (transformed) & -0.131 & -0.056 & -0.032 & -0.128 & -0.055 & -0.026 & -0.131 & -0.058 & -0.025 \\ 
  &  & QML (direct) & -0.047 & -0.008 & -0.007 & -0.047 & -0.010 & -0.003 & -0.049 & -0.011 & -0.002 \\ 
   \cline{2-12}
  &  & GMM & -0.014 & -0.005 & -0.010 & -0.017 & -0.008 & -0.004 & -0.017 & -0.007 & -0.003 \\ 
  & M3 & QML (transformed) & -0.118 & -0.045 & -0.023 & -0.117 & -0.050 & -0.022 & -0.123 & -0.053 & -0.022 \\ 
  &  & QML (direct) & -0.020 & 0.006 & 0.001 & -0.024 & -0.001 & 0.004 & -0.030 & -0.003 & 0.002 \\
  \hline
  &  & GMM & -0.075 & -0.005 & 0.001 & -0.040 & -0.012 & -0.007 & -0.015 & -0.010 & -0.003 \\ 
  & M1 & QML (transformed) & -0.300 & -0.111 & -0.047 & -0.275 & -0.120 & -0.057 & -0.263 & -0.118 & -0.054 \\ 
  &  & QML (direct) & -0.145 & -0.021 & 0.003 & -0.124 & -0.030 & -0.009 & -0.112 & -0.031 & -0.007 \\
   \cline{2-12} 
  &  & GMM & -0.059 & -0.005 & 0.001 & -0.033 & -0.012 & -0.008 & -0.010 & -0.010 & -0.004 \\ 
  $\beta_1$ & M2 & QML (transformed) & -0.284 & -0.106 & -0.044 & -0.261 & -0.115 & -0.055 & -0.249 & -0.113 & -0.052 \\ 
  &  & QML (direct) & -0.112 & -0.012 & 0.005 & -0.093 & -0.021 & -0.007 & -0.080 & -0.022 & -0.005 \\ 
   \cline{2-12}
  &  & GMM & -0.054 & -0.009 & -0.001 & -0.038 & -0.019 & -0.010 & -0.017 & -0.015 & -0.007 \\ 
  & M3 & QML (transformed) & -0.254 & -0.087 & -0.027 & -0.243 & -0.103 & -0.045 & -0.233 & -0.105 & -0.046 \\ 
  &  & QML (direct) & -0.063 & 0.014 & 0.025 & -0.053 & -0.003 & 0.006 & -0.042 & -0.007 & 0.004 \\ 
   \hline\hline
\end{tabular}
\end{scriptsize}
\end{table}

% latex table generated in R 4.3.0 by xtable 1.8-4 package
% Fri Nov 24 12:37:48 2023
\begin{table}[H]
\centering
\caption{\footnotesize RMSEs of the GMM and QML estimators across different parameter values and sample sizes} 
\label{table:rmse}
\begin{scriptsize}
\begin{tabular}{lllrrrrrrrrr}
  \hline\hline
 &  &  &  & $n=25$ &  &  & $n=49$ &  &  & $n=81$ &  \\ 
Parameter & Model & Method & $T=5$ & $T=10$ & $T=20$ & $T=5$ & $T=10$ & $T=20$ & $T=5$ & $T=10$ & $T=20$ \\ 
  \hline
   &  & GMM & 0.484 & 0.287 & 0.199 & 0.388 & 0.230 & 0.148 & 0.312 & 0.185 & 0.114 \\ 
   & M1 & QML (transformed) & 0.243 & 0.176 & 0.147 & 0.154 & 0.111 & 0.086 & 0.110 & 0.075 & 0.057 \\ 
   &  & QML (direct) & 0.177 & 0.101 & 0.066 & 0.131 & 0.078 & 0.051 & 0.102 & 0.060 & 0.040 \\ 
    \cline{2-12}
   &  & GMM & 0.473 & 0.271 & 0.185 & 0.362 & 0.217 & 0.137 & 0.291 & 0.171 & 0.106 \\ 
  $\rho$ & M2 & QML (transformed) & 0.262 & 0.182 & 0.148 & 0.164 & 0.113 & 0.085 & 0.122 & 0.078 & 0.057 \\ 
   &  & QML (direct) & 0.191 & 0.100 & 0.063 & 0.135 & 0.075 & 0.048 & 0.109 & 0.059 & 0.038 \\
    \cline{2-12} 
   &  & GMM & 0.370 & 0.223 & 0.140 & 0.231 & 0.138 & 0.085 & 0.163 & 0.099 & 0.061 \\ 
   & M3 & QML (transformed) & 0.192 & 0.146 & 0.130 & 0.093 & 0.075 & 0.061 & 0.059 & 0.046 & 0.038 \\ 
   &  & QML (direct) & 0.203 & 0.146 & 0.129 & 0.099 & 0.075 & 0.061 & 0.064 & 0.047 & 0.038 \\ 
   \hline
   &  & GMM & 0.244 & 0.102 & 0.056 & 0.161 & 0.076 & 0.038 & 0.126 & 0.055 & 0.033 \\ 
   & M1 & QML (transformed) & 0.415 & 0.176 & 0.085 & 0.400 & 0.167 & 0.077 & 0.393 & 0.160 & 0.075 \\ 
   &  & QML (direct) & 0.371 & 0.165 & 0.080 & 0.354 & 0.156 & 0.074 & 0.346 & 0.150 & 0.072 \\ 
    \cline{2-12}
   &  & GMM & 0.203 & 0.092 & 0.053 & 0.132 & 0.069 & 0.037 & 0.103 & 0.051 & 0.031 \\ 
  $\gamma$ & M2 & QML (transformed) & 0.333 & 0.142 & 0.069 & 0.322 & 0.135 & 0.063 & 0.316 & 0.130 & 0.063 \\ 
   &  & QML (direct) & 0.308 & 0.140 & 0.071 & 0.295 & 0.131 & 0.064 & 0.287 & 0.125 & 0.063 \\ 
    \cline{2-12}
   &  & GMM & 0.171 & 0.086 & 0.052 & 0.110 & 0.064 & 0.035 & 0.087 & 0.048 & 0.030 \\ 
   & M3 & QML (transformed) & 0.265 & 0.124 & 0.067 & 0.254 & 0.114 & 0.058 & 0.249 & 0.108 & 0.055 \\ 
   &  & QML (direct) & 0.254 & 0.121 & 0.067 & 0.241 & 0.111 & 0.057 & 0.235 & 0.105 & 0.055 \\ 
   \hline
   &  & GMM & 0.405 & 0.229 & 0.150 & 0.284 & 0.176 & 0.106 & 0.239 & 0.133 & 0.084 \\ 
   & M1 & QML (transformed) & 0.260 & 0.165 & 0.121 & 0.200 & 0.123 & 0.080 & 0.169 & 0.095 & 0.061 \\ 
   &  & QML (direct) & 0.242 & 0.132 & 0.086 & 0.193 & 0.105 & 0.065 & 0.165 & 0.085 & 0.053 \\ 
    \cline{2-12}
   &  & GMM & 0.398 & 0.212 & 0.142 & 0.272 & 0.169 & 0.101 & 0.239 & 0.129 & 0.080 \\ 
  $\delta$ & M2 & QML (transformed) & 0.227 & 0.146 & 0.107 & 0.153 & 0.103 & 0.070 & 0.118 & 0.075 & 0.052 \\ 
   &  & QML (direct) & 0.219 & 0.121 & 0.079 & 0.158 & 0.094 & 0.060 & 0.130 & 0.072 & 0.047 \\ 
    \cline{2-12}
   &  & GMM & 0.294 & 0.137 & 0.084 & 0.171 & 0.095 & 0.056 & 0.126 & 0.072 & 0.044 \\ 
   & M3 & QML (transformed) & 0.263 & 0.132 & 0.084 & 0.239 & 0.109 & 0.058 & 0.231 & 0.101 & 0.051 \\ 
   &  & QML (direct) & 0.261 & 0.132 & 0.084 & 0.227 & 0.106 & 0.057 & 0.214 & 0.097 & 0.050 \\ 
   \hline
   &  & GMM & 0.274 & 0.166 & 0.105 & 0.195 & 0.109 & 0.077 & 0.147 & 0.089 & 0.060 \\ 
   & M1 & QML (transformed) & 0.245 & 0.162 & 0.106 & 0.199 & 0.115 & 0.079 & 0.176 & 0.100 & 0.063 \\ 
   &  & QML (direct) & 0.261 & 0.165 & 0.103 & 0.191 & 0.108 & 0.077 & 0.151 & 0.089 & 0.059 \\ 
    \cline{2-12}
   &  & GMM & 0.272 & 0.165 & 0.105 & 0.194 & 0.109 & 0.077 & 0.146 & 0.090 & 0.060 \\ 
  $\beta_0$ & M2 & QML (transformed) & 0.242 & 0.162 & 0.106 & 0.194 & 0.114 & 0.079 & 0.172 & 0.099 & 0.063 \\ 
   &  & QML (direct) & 0.262 & 0.165 & 0.103 & 0.189 & 0.108 & 0.077 & 0.147 & 0.089 & 0.059 \\ 
    \cline{2-12}
   &  & GMM & 0.281 & 0.167 & 0.107 & 0.194 & 0.111 & 0.078 & 0.146 & 0.090 & 0.060 \\ 
   & M3 & QML (transformed) & 0.241 & 0.163 & 0.106 & 0.191 & 0.113 & 0.078 & 0.167 & 0.097 & 0.062 \\ 
   &  & QML (direct) & 0.272 & 0.171 & 0.107 & 0.192 & 0.111 & 0.079 & 0.145 & 0.089 & 0.060 \\ 
   \hline
   &  & GMM & 0.286 & 0.166 & 0.110 & 0.204 & 0.113 & 0.074 & 0.158 & 0.086 & 0.059 \\ 
   & M1 & QML (transformed) & 0.364 & 0.184 & 0.116 & 0.311 & 0.157 & 0.091 & 0.287 & 0.140 & 0.078 \\ 
   &  & QML (direct) & 0.289 & 0.161 & 0.108 & 0.218 & 0.114 & 0.073 & 0.181 & 0.089 & 0.059 \\ 
    \cline{2-12}
   &  & GMM & 0.280 & 0.165 & 0.110 & 0.200 & 0.113 & 0.074 & 0.156 & 0.086 & 0.059 \\ 
  $\beta_1$ & M2 & QML (transformed) & 0.351 & 0.182 & 0.115 & 0.299 & 0.154 & 0.089 & 0.275 & 0.136 & 0.078 \\ 
   &  & QML (direct) & 0.276 & 0.161 & 0.108 & 0.204 & 0.112 & 0.073 & 0.165 & 0.086 & 0.059 \\ 
    \cline{2-12}
   &  & GMM & 0.285 & 0.168 & 0.115 & 0.200 & 0.116 & 0.077 & 0.156 & 0.088 & 0.061 \\ 
   & M3 & QML (transformed) & 0.331 & 0.173 & 0.111 & 0.285 & 0.147 & 0.084 & 0.261 & 0.131 & 0.074 \\ 
   &  & QML (direct) & 0.275 & 0.166 & 0.116 & 0.193 & 0.113 & 0.075 & 0.155 & 0.085 & 0.060 \\ 
   \hline
\end{tabular}
\end{scriptsize}

\end{table}

%%%%%%%%%%%%%%%%%%%%%%%%
\section{Conclusion}\label{sec5}
In this paper, we considered the small and large sample properties of (i) the QMLE based on the transformation approach, (ii) the QMLE based on the direct approach, and (iii) the GMME for the estimation of a dynamic spatiotemporal ARCH model. Theoretically, the estimators based on the QML method require a large $T$ and a bias correction approach. The bias-corrected QMLE based on the transformation approach requires that $\frac{n}{T^3}\to0$, while the bias-corrected QMLE based on the direct approach requires both $\frac{n}{T^3}\to0$ and $\frac{T}{n^3}\to\infty$. The QMLE based on the transformation approach is only applicable for models with row-normalized weights matrices. The GMME does not require bias correction and is valid under both finite and large $T$ cases.  In a Monte Carlo simulation study, we compare the finite sample properties of these estimators. Our results indicate that the QMLE based on the direct approach performs relatively better for the estimation of the spatial effect ($\rho$), while the GMME performs relatively better for the estimation of the temporal effect ($\gamma$).  Overall, when both $T$ and $n$ are large enough, all estimators perform similarly and satisfactorily.  In future studies, the performance of these estimators can be investigated for a dynamic spatiotemporal ARCH model that has a multiplicative structure for the regional and time-fixed effects.

\newpage
\bibliographystyle{apalike}  
% \bibliography{ref}

%\printbibliography 

\newpage

\section*{Appendix}

\begin{figure}[H]
	\begin{center}
	    \includegraphics[width=1\textwidth]{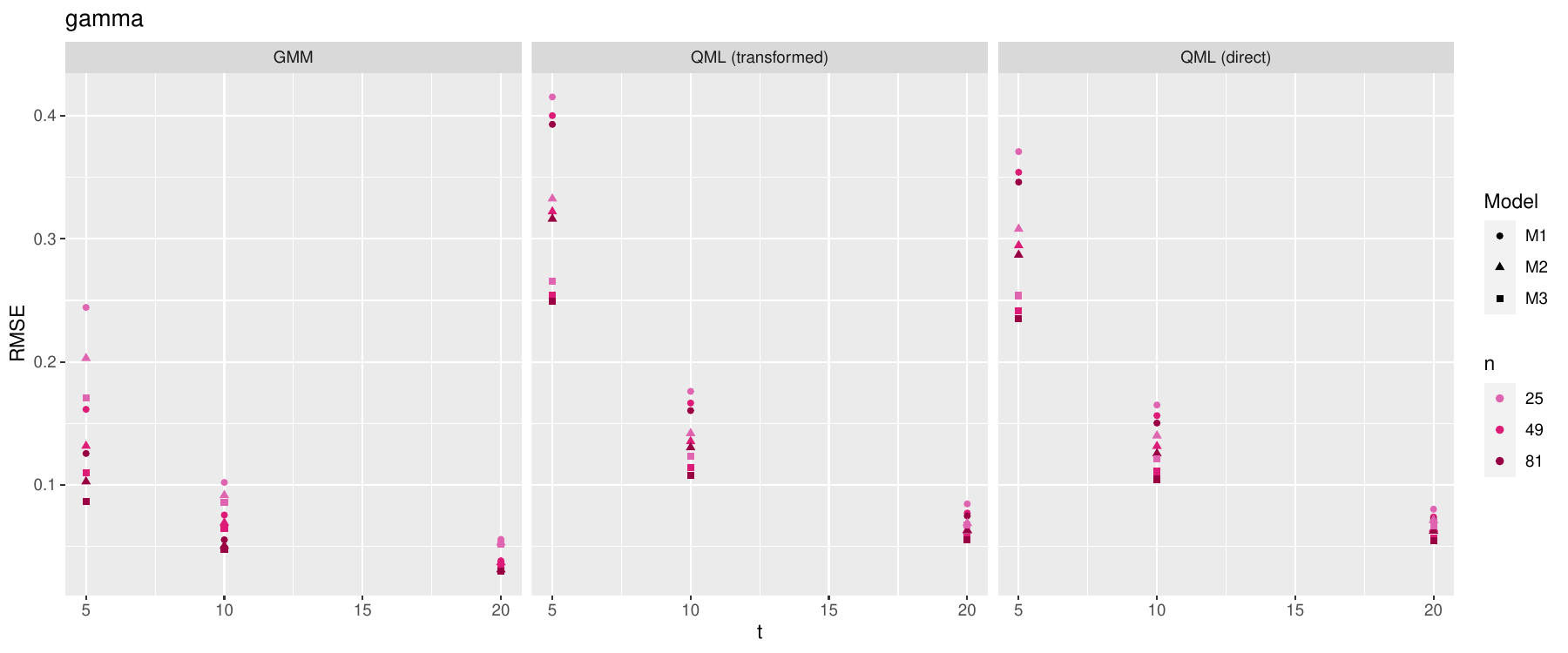}\\
	    \includegraphics[width=1\textwidth]{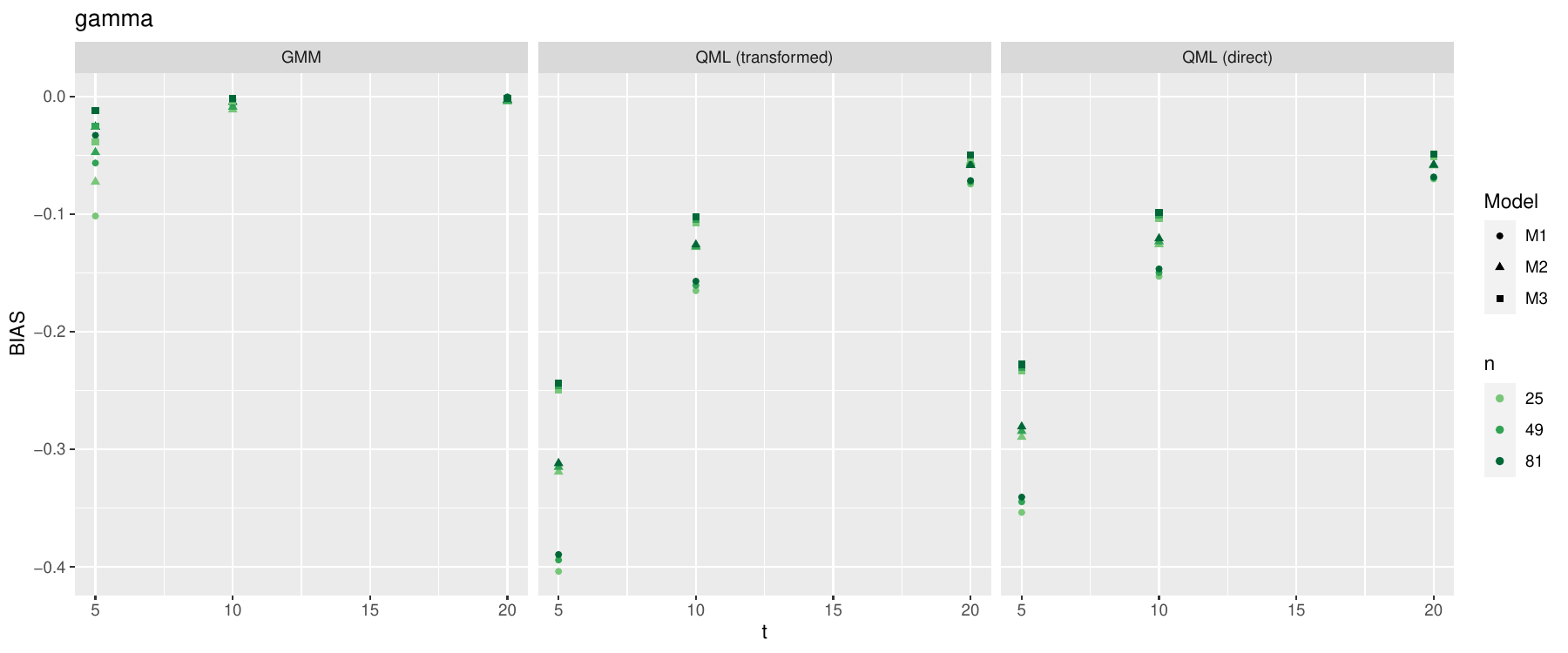}
	\end{center}
	\caption{RMSE (top panels) and average bias (bottom panels) for the estimates of the temporal autoregressive coefficient $\gamma$.}
	\label{fig:gamma}
\end{figure}

\begin{figure}[H]
	\begin{center}
	    \includegraphics[width=1\textwidth]{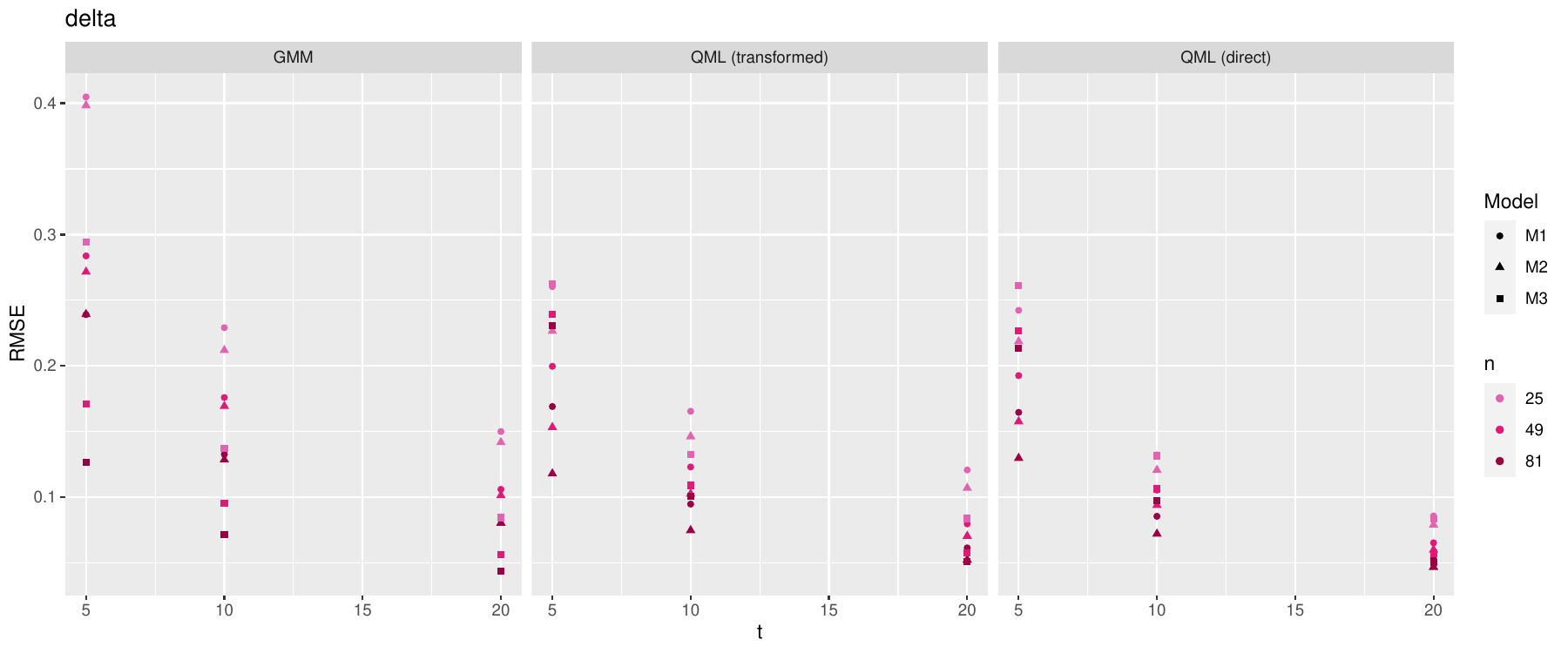}\\
	    \includegraphics[width=1\textwidth]{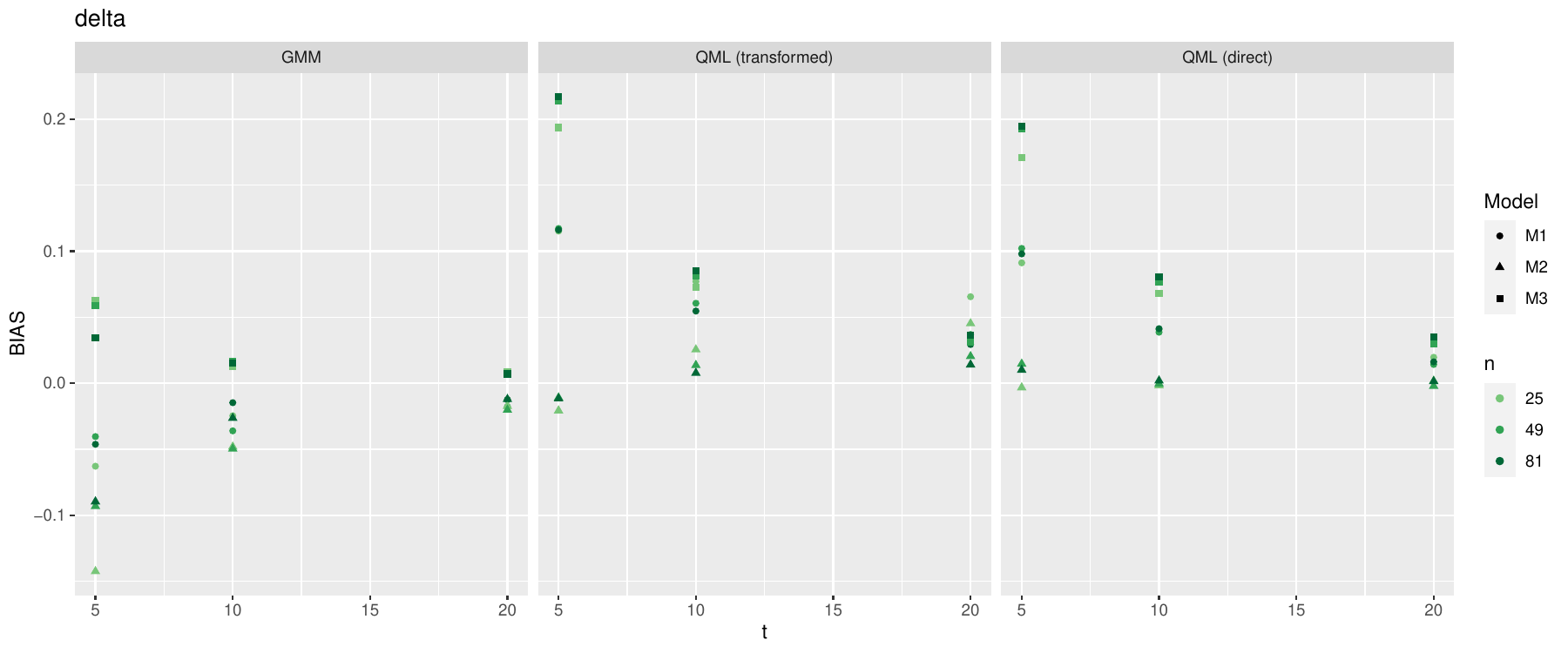}
	\end{center}
	\caption{RMSE (top panels) and average bias (bottom panels) for the estimates of the spatiotemporal autoregressive coefficient $\delta$.}
	\label{fig:delta}
\end{figure}

\begin{figure}[H]
	\begin{center}
	    \includegraphics[width=1\textwidth]{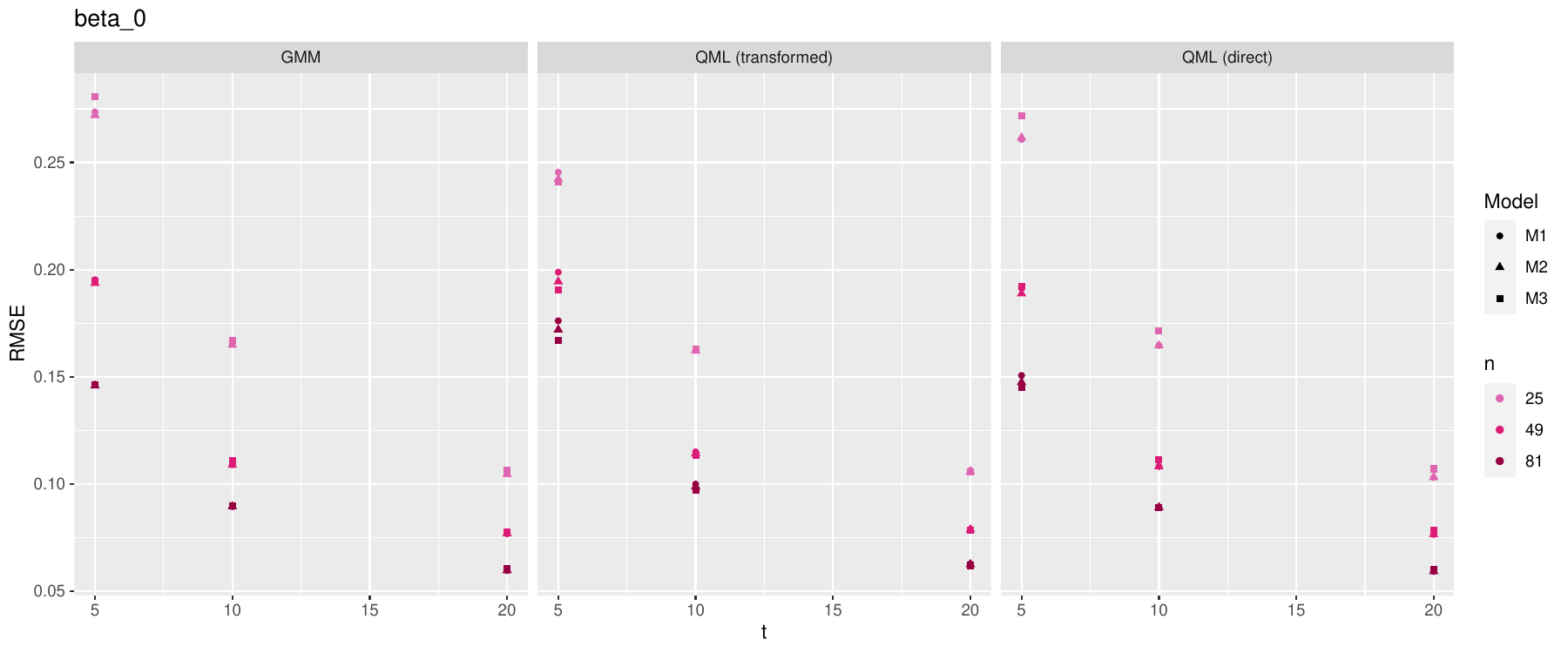}\\
	    \includegraphics[width=1\textwidth]{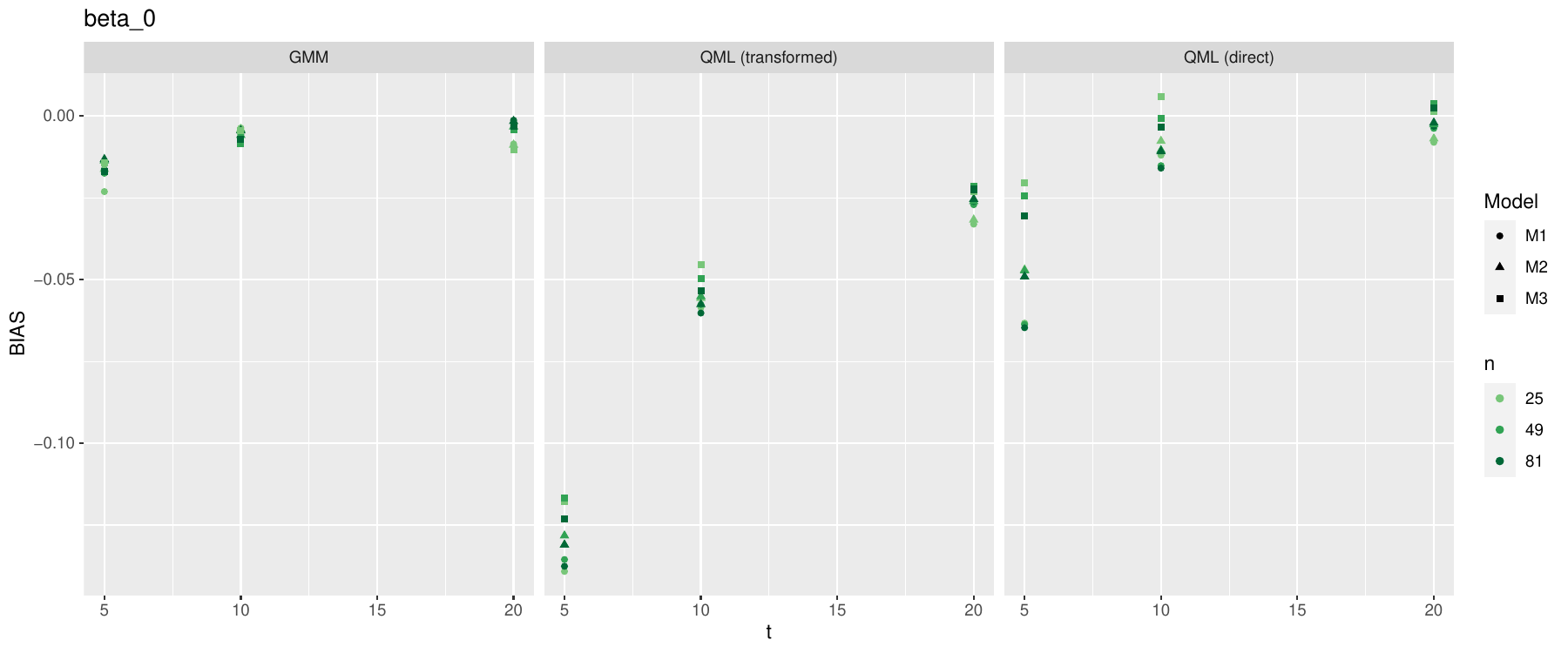}
	\end{center}
	\caption{RMSE (top panels) and average bias (bottom panels) for the estimates of the second linear regression parameter $\beta_0$.}
	\label{fig:beta_0}
\end{figure}

\begin{figure}[H]
	\begin{center}
	    \includegraphics[width=1\textwidth]{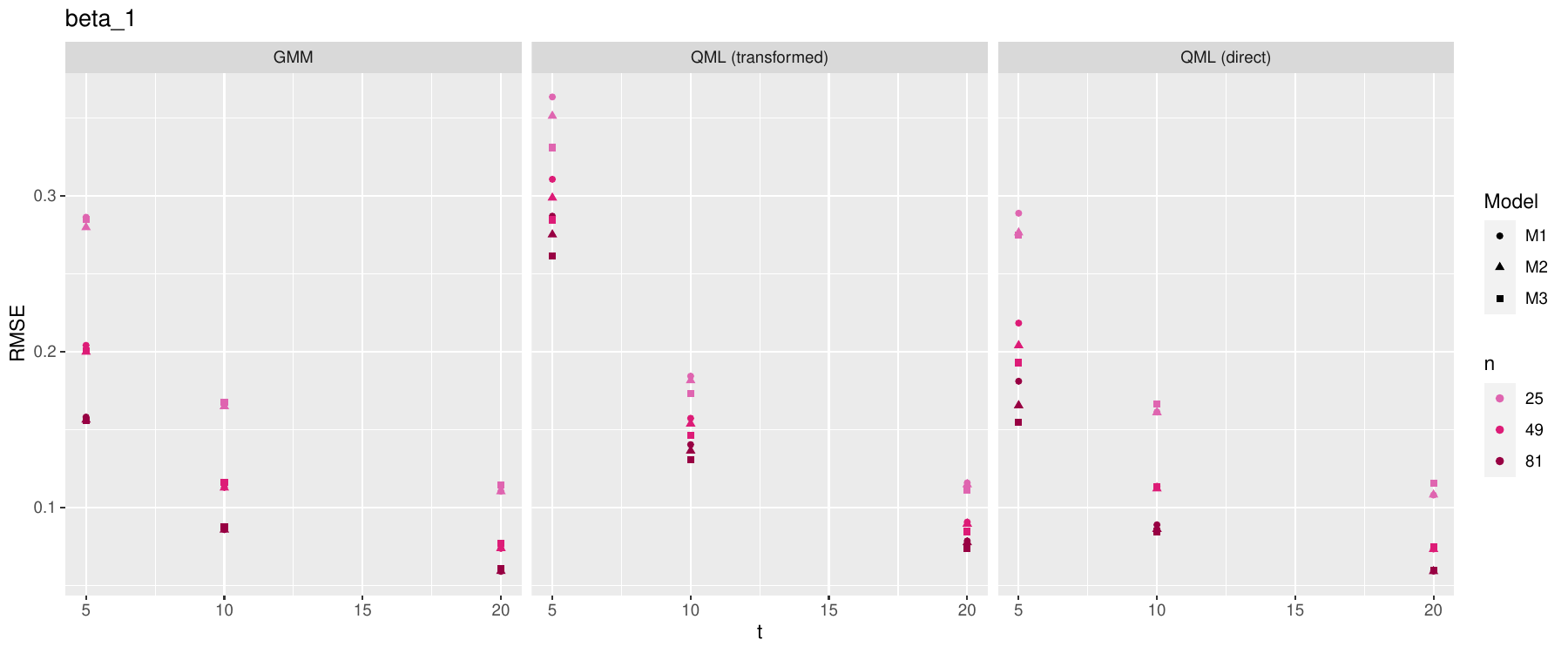}\\
	    \includegraphics[width=1\textwidth]{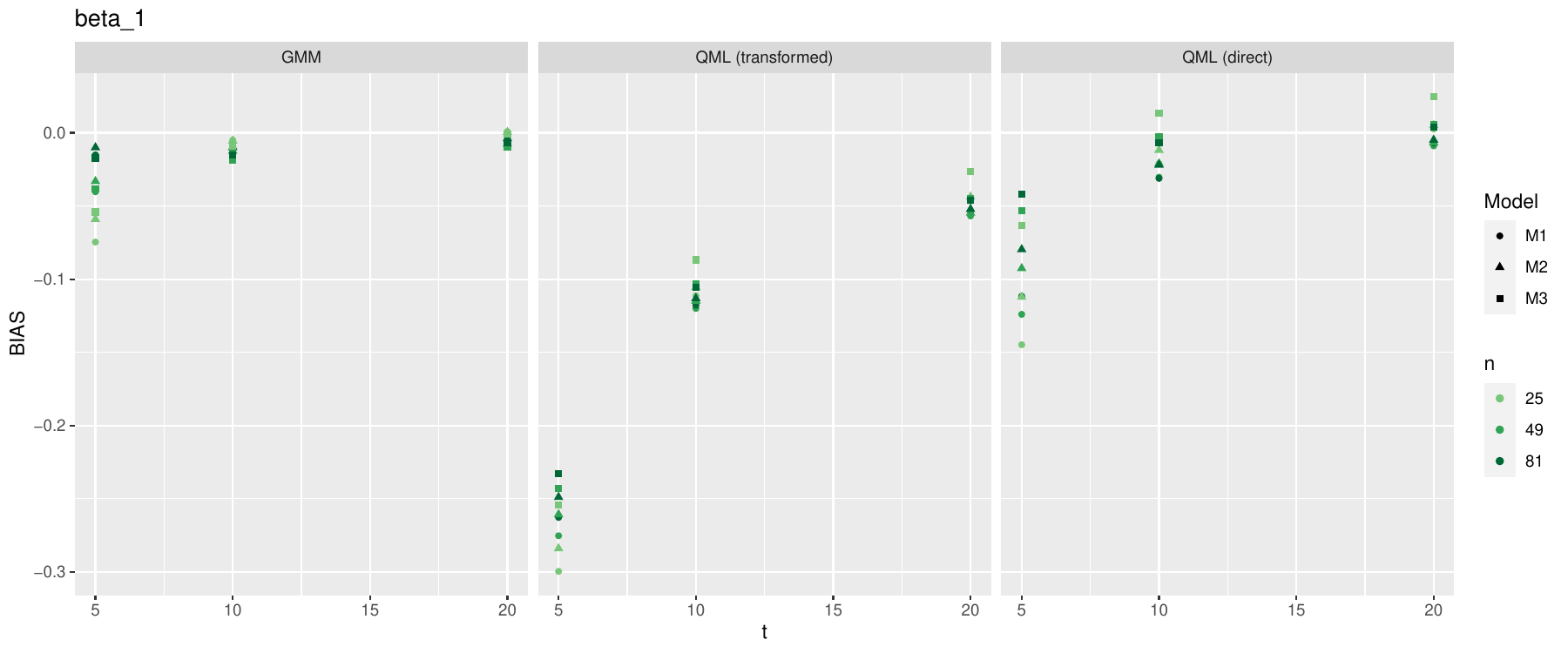}
	\end{center}
	\caption{RMSE (top panels) and average bias (bottom panels) for the estimates of the second linear regression parameter $\beta_1$.}
	\label{fig:beta_1}
\end{figure}

Below, we provide additional simulation results. Figures \ref{fig:rho} through \ref{fig:beta_1} depict graphical representations of the results provided in Tables 1 and 2 of the main manuscript. Table~\ref{table:mae} provides the mean absolute errors (MAE) for each estimator. The results in this table indicate that the MAE decreases when either $n$ or $T$ increases for all estimators. The QMLE based on the direct approach performs relatively better in estimating $\rho$, while the GMME performs relatively better in estimating $\gamma$. For all other parameters, the estimators perform similarly.

\begin{figure}[H]
	\begin{center}
	    \includegraphics[width=1\textwidth]{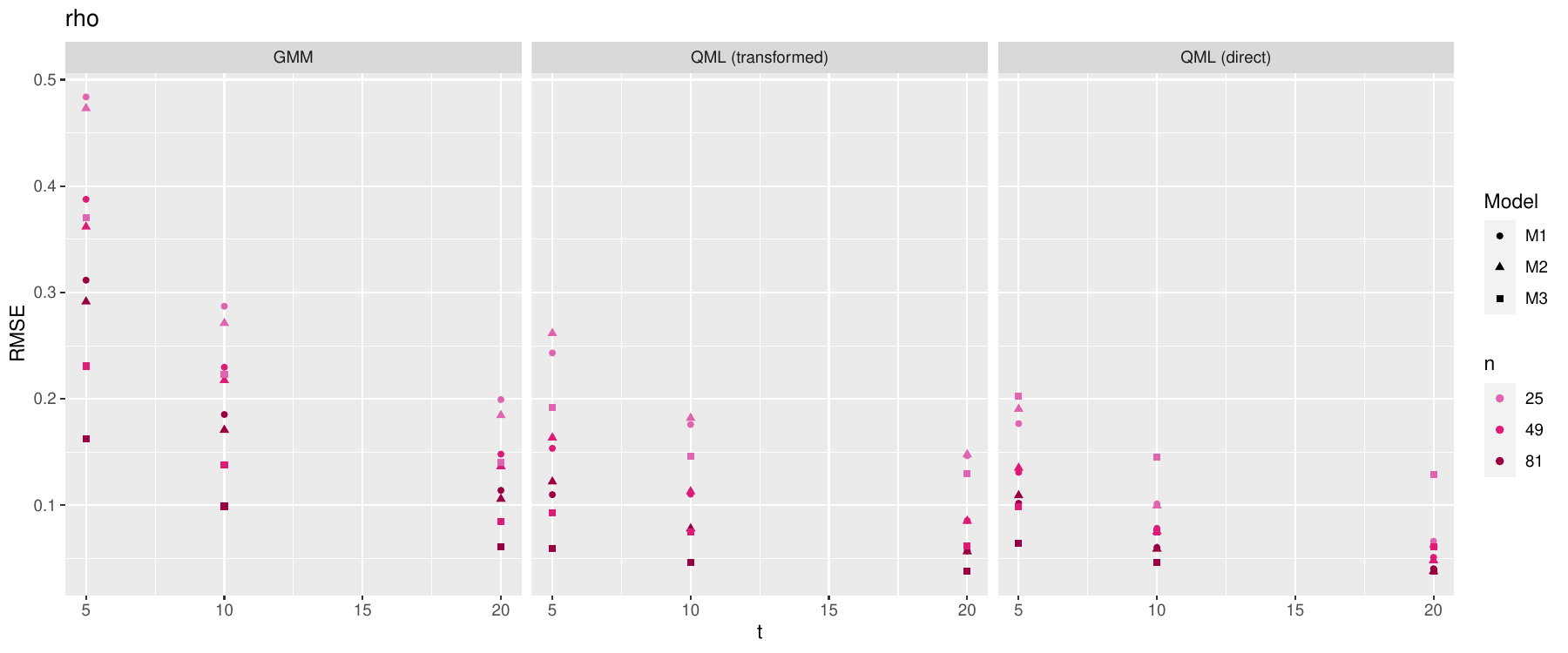}\\
	    \includegraphics[width=1\textwidth]{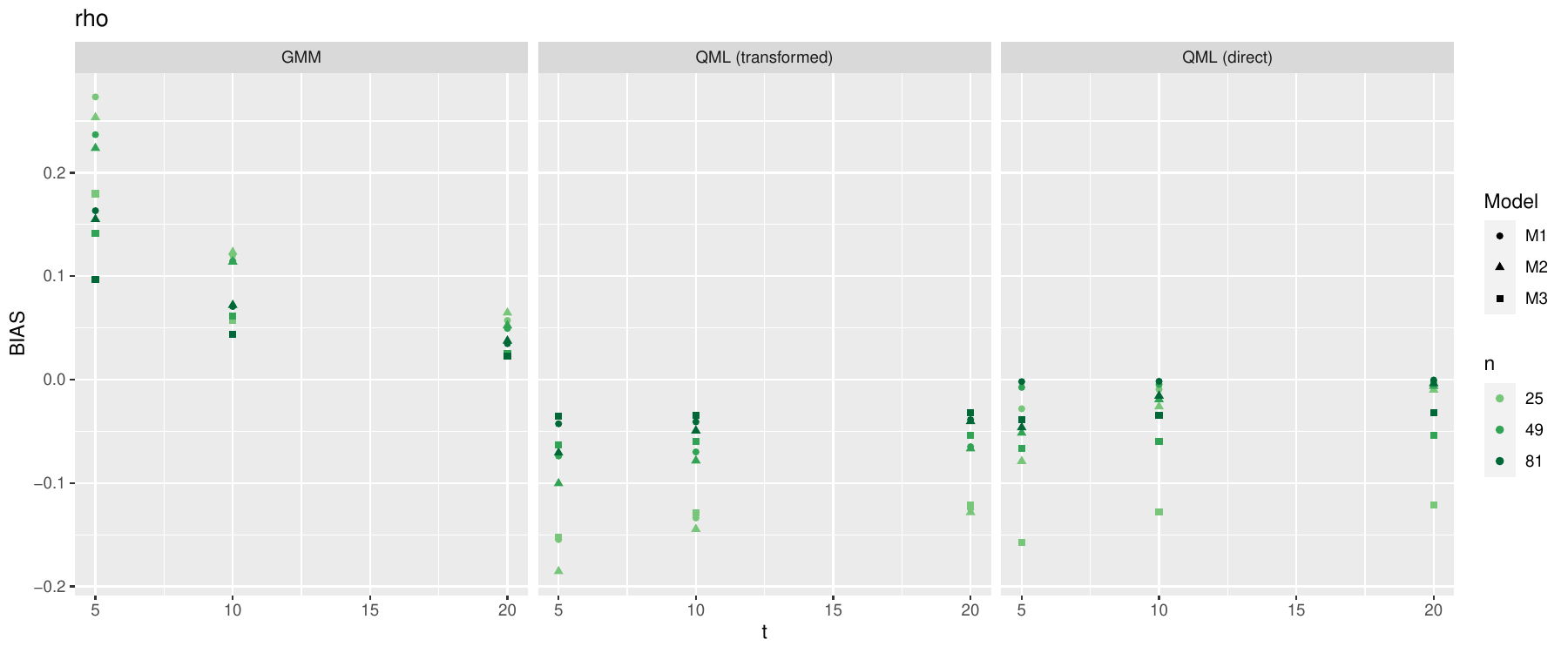}
	\end{center}
	\caption{RMSE (top panels) and average bias (bottom panels) for the estimates of the spatial autoregressive coefficient $\rho$.}
	\label{fig:rho}
\end{figure}

% latex table generated in R 4.3.0 by xtable 1.8-4 package
% Wed Nov 15 18:09:59 2023
\begin{table}[H]
\centering
\caption{\footnotesize MAEs of the GMM and QML estimators across different parameter values and sample sizes.} 
\label{table:mae}
\begin{scriptsize}
\begin{tabular}{lllrrrrrrrrr}
  \hline\hline
 &  &  &  & $n=25$ &  &  & $n=49$ &  &  & $n=81$ &  \\ 
Parameter & Model & Method & $T=5$ & $T=10$ & $T=20$ & $T=5$ & $T=10$ & $T=20$ & $T=5$ & $T=10$ & $T=20$ \\ 
  \hline
  &  & GMM & 0.405 & 0.233 & 0.160 & 0.324 & 0.190 & 0.119 & 0.259 & 0.151 & 0.092 \\ 
  & M1 & QML (transformed) & 0.193 & 0.145 & 0.128 & 0.120 & 0.088 & 0.071 & 0.087 & 0.059 & 0.046 \\ 
  &  & QML (direct) & 0.136 & 0.080 & 0.053 & 0.104 & 0.063 & 0.041 & 0.081 & 0.047 & 0.033 \\ 
  \cline{3-12}
  &  & GMM & 0.389 & 0.224 & 0.150 & 0.298 & 0.179 & 0.110 & 0.242 & 0.140 & 0.086 \\ 
  $\rho$ & M2 & QML (transformed) & 0.210 & 0.152 & 0.130 & 0.128 & 0.091 & 0.071 & 0.096 & 0.062 & 0.046 \\ 
  &  & QML (direct) & 0.144 & 0.078 & 0.050 & 0.104 & 0.060 & 0.039 & 0.085 & 0.046 & 0.031 \\ 
  \cline{3-12}
  &  & GMM & 0.291 & 0.169 & 0.109 & 0.189 & 0.110 & 0.068 & 0.136 & 0.081 & 0.050 \\ 
  & M3 & QML (transformed) & 0.157 & 0.129 & 0.122 & 0.072 & 0.062 & 0.054 & 0.046 & 0.037 & 0.033 \\ 
  &  & QML (direct) & 0.162 & 0.128 & 0.121 & 0.076 & 0.062 & 0.054 & 0.050 & 0.038 & 0.033 \\
  \hline
  &  & GMM & 0.192 & 0.079 & 0.044 & 0.126 & 0.061 & 0.031 & 0.099 & 0.044 & 0.027 \\ 
  & M1 & QML (transformed) & 0.404 & 0.165 & 0.075 & 0.394 & 0.160 & 0.072 & 0.389 & 0.157 & 0.071 \\ 
  &  & QML (direct) & 0.354 & 0.153 & 0.071 & 0.345 & 0.150 & 0.069 & 0.341 & 0.146 & 0.068 \\
  \cline{3-12} 
  &  & GMM & 0.161 & 0.072 & 0.043 & 0.105 & 0.056 & 0.029 & 0.083 & 0.040 & 0.025 \\ 
  $\gamma$ & M2 & QML (transformed) & 0.319 & 0.129 & 0.059 & 0.315 & 0.127 & 0.057 & 0.312 & 0.126 & 0.058 \\ 
  &  & QML (direct) & 0.290 & 0.127 & 0.061 & 0.284 & 0.123 & 0.058 & 0.281 & 0.121 & 0.058 \\ 
  \cline{3-12}
  &  & GMM & 0.136 & 0.068 & 0.042 & 0.087 & 0.051 & 0.028 & 0.069 & 0.038 & 0.024 \\ 
  & M3 & QML (transformed) & 0.250 & 0.109 & 0.057 & 0.245 & 0.104 & 0.051 & 0.244 & 0.102 & 0.050 \\ 
  &  & QML (direct) & 0.234 & 0.106 & 0.056 & 0.230 & 0.101 & 0.050 & 0.228 & 0.098 & 0.049 \\
  \hline
  &  & GMM & 0.308 & 0.180 & 0.119 & 0.222 & 0.135 & 0.083 & 0.185 & 0.103 & 0.067 \\ 
  & M1 & QML (transformed) & 0.208 & 0.134 & 0.096 & 0.163 & 0.096 & 0.064 & 0.138 & 0.077 & 0.049 \\ 
  &  & QML (direct) & 0.192 & 0.105 & 0.067 & 0.156 & 0.083 & 0.052 & 0.133 & 0.069 & 0.042 \\ 
  \cline{3-12}
  &  & GMM & 0.298 & 0.167 & 0.113 & 0.218 & 0.133 & 0.080 & 0.191 & 0.100 & 0.065 \\ 
  $\delta$ & M2 & QML (transformed) & 0.177 & 0.118 & 0.085 & 0.122 & 0.081 & 0.056 & 0.095 & 0.059 & 0.042 \\ 
  &  & QML (direct) & 0.172 & 0.095 & 0.062 & 0.125 & 0.074 & 0.047 & 0.104 & 0.057 & 0.037 \\ 
  \cline{3-12}
  &  & GMM & 0.231 & 0.109 & 0.068 & 0.136 & 0.074 & 0.045 & 0.102 & 0.057 & 0.034 \\ 
  & M3 & QML (transformed) & 0.218 & 0.107 & 0.066 & 0.215 & 0.091 & 0.046 & 0.217 & 0.087 & 0.042 \\ 
  &  & QML (direct) & 0.213 & 0.106 & 0.066 & 0.198 & 0.089 & 0.046 & 0.196 & 0.083 & 0.042 \\
  \hline
  &  & GMM & 0.214 & 0.135 & 0.084 & 0.157 & 0.086 & 0.061 & 0.117 & 0.071 & 0.047 \\ 
  & M1 & QML (transformed) & 0.195 & 0.130 & 0.086 & 0.164 & 0.092 & 0.064 & 0.149 & 0.081 & 0.050 \\ 
  &  & QML (direct) & 0.206 & 0.134 & 0.083 & 0.154 & 0.085 & 0.061 & 0.121 & 0.071 & 0.047 \\ 
  \cline{3-12}
  &  & GMM & 0.214 & 0.134 & 0.084 & 0.155 & 0.086 & 0.061 & 0.117 & 0.071 & 0.047 \\ 
  $\beta_0$ & M2 & QML (transformed) & 0.192 & 0.131 & 0.086 & 0.159 & 0.091 & 0.063 & 0.144 & 0.080 & 0.050 \\ 
  &  & QML (direct) & 0.206 & 0.134 & 0.083 & 0.152 & 0.085 & 0.061 & 0.119 & 0.071 & 0.047 \\
  \cline{3-12} 
  &  & GMM & 0.219 & 0.135 & 0.085 & 0.156 & 0.088 & 0.062 & 0.118 & 0.072 & 0.048 \\ 
  & M3 & QML (transformed) & 0.190 & 0.131 & 0.085 & 0.155 & 0.090 & 0.063 & 0.139 & 0.078 & 0.049 \\ 
  &  & QML (direct) & 0.213 & 0.139 & 0.086 & 0.154 & 0.088 & 0.062 & 0.117 & 0.071 & 0.048 \\
  \hline
  &  & GMM & 0.228 & 0.133 & 0.089 & 0.164 & 0.091 & 0.059 & 0.126 & 0.069 & 0.047 \\ 
  & M1 & QML (transformed) & 0.315 & 0.150 & 0.093 & 0.278 & 0.132 & 0.073 & 0.264 & 0.122 & 0.064 \\ 
  &  & QML (direct) & 0.234 & 0.129 & 0.087 & 0.175 & 0.092 & 0.058 & 0.148 & 0.071 & 0.048 \\ 
  \cline{3-12}
  &  & GMM & 0.223 & 0.132 & 0.089 & 0.161 & 0.091 & 0.059 & 0.125 & 0.069 & 0.047 \\ 
  $\beta_1$ & M2 & QML (transformed) & 0.302 & 0.147 & 0.093 & 0.264 & 0.128 & 0.071 & 0.251 & 0.118 & 0.063 \\ 
  &  & QML (direct) & 0.222 & 0.129 & 0.087 & 0.163 & 0.090 & 0.058 & 0.134 & 0.069 & 0.047 \\ 
  \cline{3-12}
  &  & GMM & 0.227 & 0.134 & 0.092 & 0.160 & 0.093 & 0.061 & 0.124 & 0.071 & 0.049 \\ 
  & M3 & QML (transformed) & 0.281 & 0.139 & 0.090 & 0.248 & 0.120 & 0.067 & 0.235 & 0.112 & 0.060 \\ 
  &  & QML (direct) & 0.216 & 0.133 & 0.093 & 0.153 & 0.091 & 0.060 & 0.125 & 0.068 & 0.048 \\ 
   \hline\hline
\end{tabular}
\end{scriptsize}
\end{table}

\end{document}